\documentclass{aa}  

\usepackage{graphicx}
\usepackage{multirow}
\usepackage{amsmath}	
\usepackage{longtable}
\usepackage{threeparttablex}
\usepackage{footnote}
\usepackage{xcolor}
\usepackage{hyperref}
\usepackage{booktabs}
\usepackage{multirow}
\usepackage{placeins}

\hypersetup{colorlinks=true,citecolor=blue, linkcolor=blue}
\usepackage{txfonts}
\newcommand{\funit}{~erg cm$^{-2}$ s$^{-1}$} 
\newcommand{\ergs}{~erg s$^{-1}$}

\newcommand{\cmsq}{~cm$^{-2}$ }   

\newcommand{\chisq}{$\rm{\chi^{2}}$ }

\def\xmm{{\it XMM-Newton}} 
\def\chandra{{\it Chandra}}

\begin{document}

   \title{New X-ray Supernova Remnants in NGC 7793}

   \subtitle{}

   \author{
    Kopsacheili M.\inst{1,2} \and  
    Anastasopoulou K.\inst{3} \and  
    Nanda R.\inst{1,2} \and  
    Gutierrez C. P.\inst{2,1} \and  
    Galbany L.\inst{1,2}  
}

    \institute{Institute of Space Sciences (ICE, CSIC), Campus UAB, Carrer de Can Magrans, s/n, E-08193 Barcelona, Spain \\
    \email{kopsacheili@ice.csic.es}
    \and Institut d'Estudis Espacials de Catalunya (IEEC), 08860  Castelldefels (Barcelona), Spain  
    \and Center for Astrophysics $|$ Harvard \& Smithsonian, 60 Garden Street, Cambridge, MA 02138, USA 
             }

   \date{received: March 7, 2025; accepted: June 5, 2025}

 
  \abstract
   {This work focuses on the detection of X-ray Supernova Remnants (SNRs) in the galaxy NGC 7793 and the study of their properties.}
   {X-ray SNRs in galaxies beyond the Local Group are rare, mainly due to the limited sensitivity of current X-ray instruments. Additionally, their identification requires an optical counterpart, making incomplete optical identification methods an extra challenge. Detecting X-ray SNRs in other galaxies is crucial for understanding their feedback in different evolutionary phases and gaining insights into their local interstellar medium (ISM). In NGC 7793, only one X-ray SNR was previously known, while a recent study reported nearly 240 optical SNRs. The discovery of a new, larger optical SNR sample motivated a re-examination of the X-ray SNR population by comparing optical SNRs with X-ray sources.}
   {To identify X-ray SNRs, we utilized \chandra’s spatial resolution and analyzed all available archival data of NGC 7793, totaling 229.9 ks over 19 years. After data reduction, we performed source detection and analysis, searching for X-ray sources coinciding with optical SNRs. We also used \xmm{} (1.1 Ms combined EPIC MOS) for spectral analysis of the confirmed and candidate SNRs.}
   {We detected 58 X-ray sources  down to an observed luminosity of $\rm \sim 1.5\times 10^{36}\, erg\, s^{-1}$. Among them, five X-ray counterparts to optical SNRs were identified, all presenting soft emission (<1.2 keV) with no short- or long-term variability. One corresponds to the previously known X-ray SNR, while four are newly detected. Spectral modeling of two SNRs shows thermal spectra exceeding 2.5 million K, with strong O\,VII, O\,VIII, and Ne\,IX emission lines. A correlation between density, X-ray luminosity, and source softness was observed. We also report X-ray emission from supernova 2008bk, refining its position, and suggest two candidate X-ray SNRs with soft, non-variable spectra, one resembling the identified X-ray SNRs.}
   {}


   \maketitle
%

\section{Introduction}
Supernova Remnants (SNRs) are very important ingredients of galaxies. They enrich the Interstellar Medium (ISM) with heavy elements and they depose to it large amounts of  energy that heat and shape it. In addition, they can trigger star formation when their shock waves compress nearby molecular clouds. Their systematic study can give information on their feedback to the ISM. In case of core-collapse SNRs, they trace the massive-star formation rate of a galaxy, since those SNRs depict massive star's endpoint life (e.g. \citealt{1997ARA&A..35..309F}).

The evolution of SNRs can be described by four consecutive phases: free expansion, adiabatic or Sedov-Taylor phase, radiative phase and fade-out phase (e.g. \citealt{1975ApJ...200..698C}; \citealt{1988ApJ...334..252C}). We usually observe SNRs in the Sedov-Taylor and radiative phases, as they last longer and during this time  the SNRs' energetics reach their peak. In the first two phases, SNRs mostly emit in X-rays, where their fast shocks heat the ISM to temperatures of $\sim 10^7\, \rm K$, producing thermal emission. However, non-thermal emission can be produced from the central object, for example a pulsar or pulsar wind nebula (PWN) or, more rarely, in the SNR's shell, because of relativistic particles accelerated by the shock wave. The optical emission is usually expected during the radiative phase (or at the end of the Sedov-Taylor phase), when the shock slows down and the SNR's temperature decreases. However, there are SNRs that are observed in both optical and X-rays, which probably indicates that they evolve in environments with high density variations. The portion of the SNR's shell that encounters denser ISM cools down and can produce optical emission lines, while the rest keeps moving with higher velocities, emitting in X-rays. There is another category of SNRs where optical (and/or radio) and thermal X-ray emission coexist. This is the mixed-morphology (or thermal composite) SNRs where the optical/radio emission comes from the shell and the X-ray emission from its central region. This category is not well understood yet, however there are some explanations for its nature: (a) evaporating cloud in the interior of the SNRs may emit soft X-rays because of the evaporated material (\citealt{1994ApJ...430..757R}; \citealt{1991ApJ...373..543W}); (b) the SNR's shell collide with the density wall of a pre-existing cavity that has been formed by stellar winds of the progenitor star. Then a reflected shock reheats the SNR's interior producing soft X-ray emission (\citealt{2005ApJ...630..892D}; \citealt{2008ApJ...676.1040C}; \citealt{2015ApJ...799..103Z}; \citealt{2022MNRAS.512.1658B}); (c) the density in the interior of the SNR can be increased by thermal conduction and so does the temperature. The hot plasma emits in X-rays (\citealt{1999ApJ...524..179C}; \citealt{1999ApJ...524..192S}). The most recent model was presented by \citet{2024MNRAS.531.5109C}, who proposed an SNR evolving in a cavity created by stellar winds of a massive-star progenitor. The forward shock encounters the density wall of the cavity, and then a reflected shock reheats the supernova (SN) ejecta and the red supergiant bubble created during the last stage of the progenitor's life. There is another category of SNRs,  the (non-thermal) composite SNRs,  that may present an optical/radio shell and non-thermal X-ray emission powered by a central object, such as a neutron star.   
A multi-wavelength study of extragalactic SNRs is of high significance in order to understand their feedback to the ISM, and how ISM properties affect their evolution. The last decades a lot of surveys have been published on Galactic SNRs and SNRs in Magellanic Clouds (MCs) in different wavelengths (e.g: \citealt{2006ApJ...642..260S}; \citealt{2008A&A...485...63F}; \citealt{2013ApJ...772..134M}; \citealt{2019A&A...631A.127M}; \citealt{2020ApJ...898L..51W}; \citealt{2021ApJ...920...90F}; \citealt{2022MNRAS.512.1658B}; \citealt{2021MNRAS.507..971C},\citeyear{2022MNRAS.513L..83C}; \citealt{2022ApJ...932...26T}; \citealt{2022MNRAS.tmp.1558P}; \citealt{2022MNRAS.514..728A}; \citealt{2024A&A...692A.237Z}). Their proximity allows the detailed investigation of their physical properties, morphological and kinematic characteristics, their interaction with their ambient medium, and their progenitor. Especially young SNRs, with prominent emission in X-rays reveal information on the yields of the SN explosion, and on the circumstellar medium (CSM) or the shaped by the progenitor ISM (e.g. \citealt{2012A&ARv..20...49V}).

On the other hand, extragalactic SNRs provide the opportunity to study larger samples in various galactic environments, e.g. in galaxies with different morphology, metallicity, star formation rate (SFR). Optical studies have historically provided the highest number of extragalactic SNRs. Theoretical and observational studies in optical, such as \citet{2013MNRAS.429..189L}; \citet{2019ApJ...875...85L}; \citet{2021MNRAS.502.1386C}; \citet{2021MNRAS.507.6020K}; \citet{2022MNRAS.514.3260K}, have revealed properties of SNR populations such density, shock velocities and their connection to the ambient ISM. On the other hand,  in other wavelengths the sample size is limited, primarily due to the sensitivity constraints of the instruments, especially in galaxies out of the Local Group (e.g. Fig. 13 in \citealt{2013MNRAS.429..189L} and Fig. 7 in \citealt{2017ApJS..230....2B}). Other studies have been dedicated to the simultaneous exploration of optical and X-ray or radio properties, such as \citet{2007AJ....133.1361P}; \citet{2010ApJ...725..842L}; \citet{2011AAS...21725633P}; \citet{2021ApJ...908...80W}.

In this work, we use all the available archival {\it{Chandra}} data of NGC 7793 in order to seek for new X-ray SNRs. NGC 7793 is a flocculent, spiral, almost face on galaxy, member of the Sculptor Group
\citep{1988AJ.....95.1025P}, at a distance of 3.7 Mpc. It presents an SFR of $\rm 0.51 M\odot \, yr^{-1}$, based on the exctinction corrected H$\rm \alpha$ luminosity \citep{2009ApJ...706..599L}. Despite the large number of optical SNRs detected in galaxies beyond the Local Group, only a few X-ray SNR counterparts have been identified (e.g., \citealt{2007AJ....133.1361P}; \citealt{2010ApJ...725..842L}). NGC 7793 is an excellent candidate for expanding the X-ray SNR sample, as it hosts a large number of optical SNRs. While previous X-ray studies have reported a single X-ray SNR in this galaxy, the availability of new and larger optical SNR catalogs motivated us to revisit the search for additional X-ray SNRs.

In optical, \citet{Blair1997} detected and spectroscopically confirmed 27 SNRs, based on the [\ion{S}{II}]/H$\rm \alpha$ > 0.4 criterion.  Two of them have been also confirmed by \citet{2020AA...635A.134D}. Later on, \citet{2021MNRAS.507.6020K} presented 55 candidate optical SNRs, 29 of which are new identifications, and more recently \citet{2024MNRAS.530.1078K} identified $\sim$ 238 SNRs, where $\sim$ 225 of which are new identification, based on multi-line diagnostics developed by \citet{2020MNRAS.491..889K}. In radio, \citet{2002ApJ...565..966P} presented 7 radio SNRs, 2 of which coincide with optical SNRs, while the rest of them were new identifications. More recently, \citet{2014SerAJ.189...15G} presented a catalogue of 14 radio SNRs  that includes 5 of the 7 aforementioned radio SNRs. Two of them coincide with optical SNRs from the work of \citet{Blair1997}. In X-rays \citet{2011AJ....142...20P} identified 1 SNR, counterpart to the optical SNR S11 in \citet{Blair1997}.

The structure of the paper is the following: In \S \ref{sec:obs_red} the X-ray observation details are presented, as well as the data reduction, the detection and analysis, and the hardness ratio calculation.
In \S \ref{sec:results} we present our results, and in \S \ref{sec:discussion}, we compare them to other studies. In this section we also examine correlations between X-ray and optical properties. Finally, in \S \ref{sec:conclusions} we summarize our results and draw our conclusions.

\section{Observations and Data Analysis} \label{sec:obs_red}
\subsection{Chandra data}
In this work we used {\textit{Chandra}} \citep{2000SPIE.4012....2W} archival data obtained by the ACIS-S camera \citep{1997AAS...190.3404G} between 2003 and 2020, for a total exposure time of 229.9 ks. In \autoref{table:chandra_info} we present the details of each observation. The data analysis was performed using the \texttt{CIAO} software version 4.14 and \texttt{CALDB} version 4.9.7 
\citep{fruscione06}.

\begin{table}
\tiny
	\centering
		\caption{Archival \chandra{} observations of NGC\,7793 with ACIS-S}
	\begin{tabular}{llcccl} 
		\hline
		OBS ID & Exp.  & PI & RA & DEC & Start Date\\
		       & (ks)  &    & (J2000) & (J2000) & \\
		\hline
		3954 & 48.94 & Pannuti    & 23:57:49.8  & -32:35:29.5 & 2003-09-06\\
		14231 & 58.84 & Soria     & 23:57:59.9	& -32:33:20.9 & 2011-08-13\\
		13439 & 57.77 & Soria     & 23:57:59.9	& -32:33:20.9 & 2011-12-25\\
		14378 & 24.71 & Soria     & 23:57:59.9	& -32:33:20.9 & 2011-12-30\\
		23266 & 29.69 & Walton    & 23:57:51.0  & -32:37:26.6 & 2020-06-04\\
            27481 & 9.95  & Brightman & 23:57:49.9  & -32:35:27.7 & 2022-10-27\\
		\hline
	\end{tabular}
	\label{table:chandra_info}
	
\end{table}

We followed  the standard {\it{Chandra}} analysis tools, and we started by running the \textit{chandra\_repro} script for each observation, that creates a new bad pixel file, and a level-2 event file. We also checked for background flares and we found that the background was constant throughout all the observations.

Before the detection process, aiming to increase the signal to noise ratio (S/N) of the X-ray sources of the galaxy, the observations were combined using the \textit{merge\_obs} tool which first reprojects all the observations to a common target point and then creates a merged Level-2 event file. We note that the merged observations are used only for the source detection and no other further analysis. During this process, for each OBSID, images and exposure maps were produced in 4 bands: broad (0.5-7.0 keV), soft (0.5-1.2 keV), medium (1.2-2.0 keV), and hard (2.0-7.0 keV).

\subsubsection{Source detection and fluxes} \label{sec:detection}
For the detection we used the \textit{wavedetect} tool \citep{2002ApJS..138..185F}  and we searched for sources in the broad, hard, medium, and soft images of the individual observations as well as of the merged one on scales of 1, 2, 4, 8, and 16 pixels. We then compared the outputs from all detections across bands and observations, and if a source appeared more than once, we retained it only once in the final source list.

In order to infer the count rates, for each detected source the PSF (Point Spread Function) that includes the 90\% of the light in the broad band, was calculated using the \textit{psfsize\_srcs} tool. For the background, an annulus was defined with inner radius 2 pixels larger than the source aperture, and outer radius double the radius of the source aperture. This way we avoid contamination from the wings of  the source's PSF and  we ensure good statistics of the counts. We visually inspected the source and background regions to avoid contamination from nearby sources. The source counts were extracted with the \textit{dmextract} tool in the individual observations of the broad, soft, medium, and hard bands. The errors correspond to a 1$\sigma$ Gaussian distribution.
Since supernova remnants are expected to be predominantly soft and not variable over the time frame of the \chandra{} observations, fluxes were calculated using the OBSID with the highest signal-to-noise ratio in the soft band. Combined observations were not used since the \chandra{} effective area, particularly at low energies, has changed over time due to contamination build-up on the ACIS optical blocking filter \citep[e.g.][]{odell15}. This can introduce systematic errors in the derived fluxes when combining data from different epochs. 

Fluxes were calculated only for a subset of sources (detailed in \S\ref{sec:correlations}) for which we have indications of their nature, allowing us to apply appropriate spectral models. To do this, we used the {\it{Chandra Proposal Planning Toolkit} (PIMMS\footnote{https://cxc.harvard.edu/toolkit/pimms.jsp})}. This tool allows for the calculation of the flux of
a source given its count rate in cases where the source count statistics do not allow for spectral extraction and modeling. The model and parameters used are detailed in that section.\footnote{
A similar analysis could, in principle, be performed using information from the Chandra Source Catalog \citep{evans10}. However, one of the observations is not included in the latest CSC release (v2.1). In order to ensure consistency and to apply SNR-appropriate models, we reanalyzed the data.

}

\subsubsection{Calculation of Hardness Ratios  and color-color diagrams} \label{sec:HR}
Extragalactic SNRs usually are faint objects in X-rays. This does not allow for any meaningful spectral modeling and exploration of their spectral properties. In order to estimate the spectral properties of the X-ray sources we calculated the hardness ratio of each detected source which is usually the ratio of the counts of two bands or a monotonic function of it \citep{2006ApJ...652..610P}. For the calculation of the hardness ratios we used the BEHR code (Bayesian Estimation of Hardness Ratios; \citealt{2006ApJ...652..610P}). This tool assumes Poisson distributions, a proper assumption in our case where most of the sources have low counts,  and it evaluates the posterior probability distribution of the colors, providing reliable estimations and correct confidence limits, even for very low counts. In addition, it can estimate upper and lower limits in cases where a source is not detectable in one of the passbands. 

We ran the BEHR code for the observation that presents the higher S/N in the soft band for each source, since we expect SNRs to present higher emission at lower energies. The hardness ratios that we use are: for the bands soft-hard: $\rm log_{10}(S/H)$; soft-medium: $\rm log_{10}(S/M)$, and medium-hard: $\rm log_{10}(M/H)$, where soft, medium and hard  refer to the net counts with energy ranges: 0.5$-$1.2 keV,  1.2$-$2.0 keV, and 2.0$-$7.0 keV respectively. The uncertainties of the hardness ratios were estimated at the 68\% confidence level.

\section{Results} \label{sec:results}
\begin{figure*}[htbp]
\centering
\includegraphics[width=0.95\textwidth]{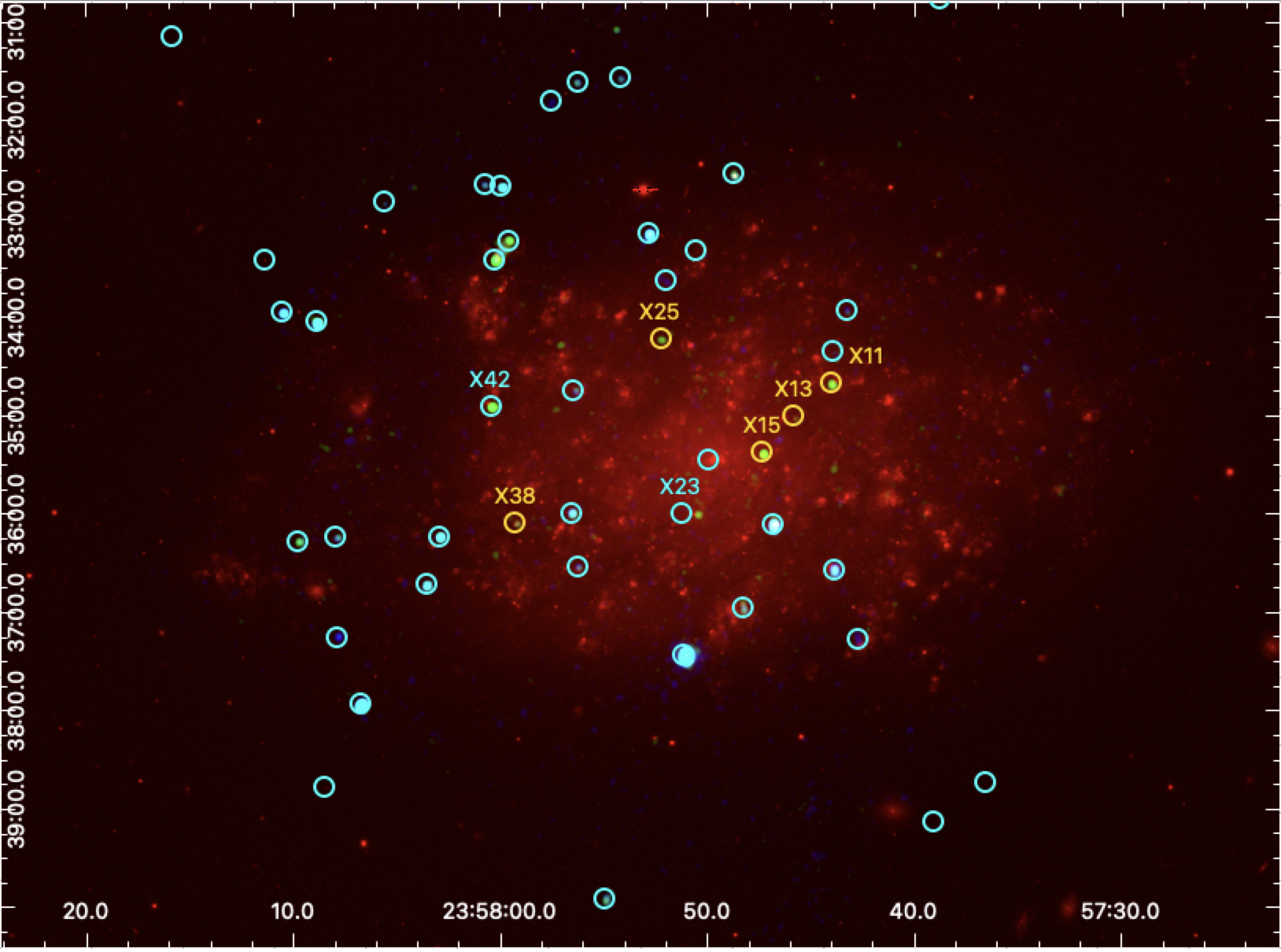}
\caption{\label{fig:7793_rgb} Composite X-ray optical image of NGC 7793 consisting of: H$\rm \alpha$ + [N II] (red), soft X-ray (0.5 - 1.2 keV; green), and medium + hard X-ray (1.2 - 7.0 keV; blue) of the OBSID 3954. All circles indicate the X-ray detected sources. The orange circles show the X-ray sources that coincide with optical SNRs that have been identified in \citet{2021MNRAS.507.6020K} and \citet{2024MNRAS.530.1078K}.}
\end{figure*}

We detected 58 X-ray sources which are listed in \autoref{table:source_info}, along with their coordinates, the net count rates, and the S/N ratios in the soft and broad band. 
In \autoref{table:fluxes} of \S \autoref{sec:comparison} we present the selected OBSID of each source (i.e. the OBSID for which the S/N ratio in the soft band is higher) and the colors $\rm log_{10}(S/H)$, $\rm log_{10}(S/M)$, and $\rm log_{10}(M/H)$. 
In \autoref{fig:7793_rgb} we show the composite X-ray optical image of NGC 7793 consisting of: H$\rm \alpha$ (red), soft X-ray (0.5 - 1.2 keV; green), and medium + hard X-ray (1.2 - 7.0 keV; blue). All circles indicate the X-ray detected sources. The orange circles show the X-ray sources that coincide with optical SNRs that have been identified in \citet{2021MNRAS.507.6020K} and \citet{2024MNRAS.530.1078K} and are the main focus of this study.

\section{Discussion} \label{sec:discussion}
\subsection{Comparison of X-ray sources with other surveys} \label{sec:comparison}
Of the 58 detected sources, 30 are reported for the first time in X-rays: X1, X2, X3, X4, X5, X6, X7, X8, X10, X16, X17, X20, X23, X24, X27, X29, X30, X31, X32, X35, X36, X37, X38, X44, X47, X48, X49, X51, and X55. Among these, 19 sources have S/N > 3, six sources have 2 < S/N < 3, and five sources have 1 < S/N < 2 (S/N in the broad band). From the remaining sources those that are unrelated to SNRs are compared with other surveys in \S \autoref{sec:comparison_ap} (and \autoref{table:fluxes}). Subsequently, we talk about the detection of X-ray emission of the supernova 2008bk and then about sources that coincide with optical SNRs or could be associated with them.

\begin{figure*}[htbp]
\minipage{0.9\textwidth}
\includegraphics[width=0.32\linewidth]{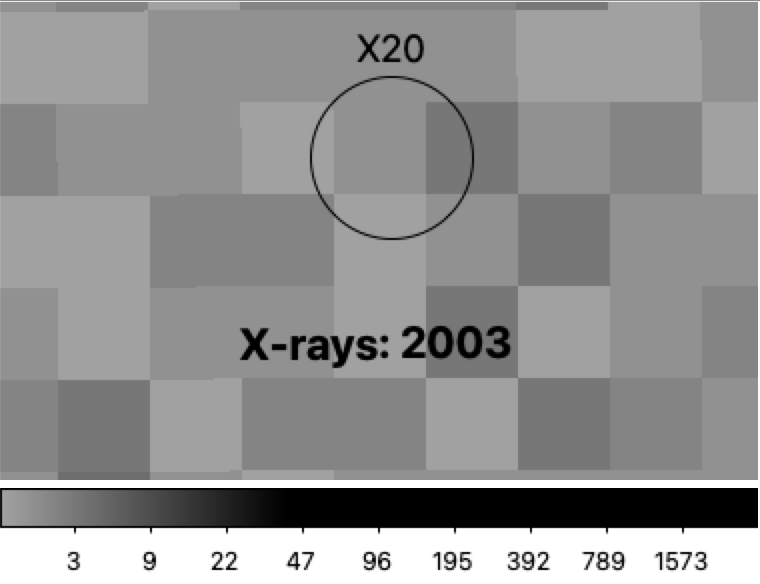}
\hfil
 \includegraphics[width=0.32\linewidth]{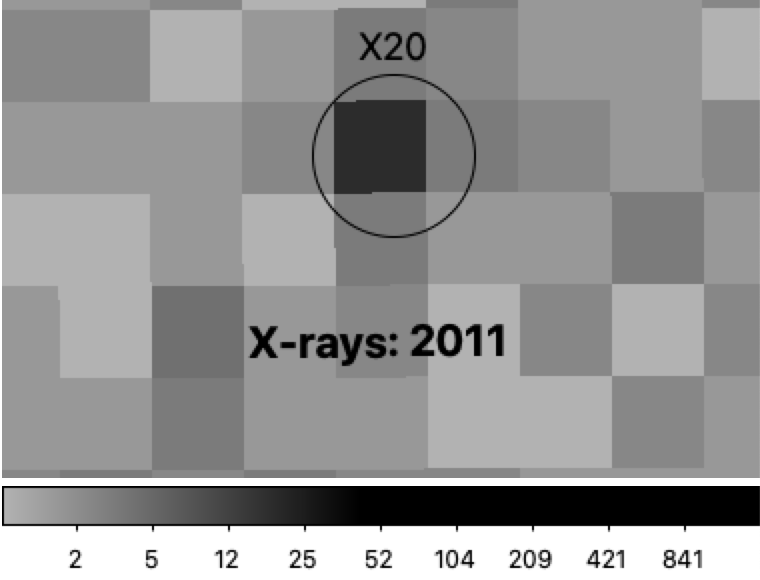}
\hfil
\includegraphics[width=0.32\linewidth]{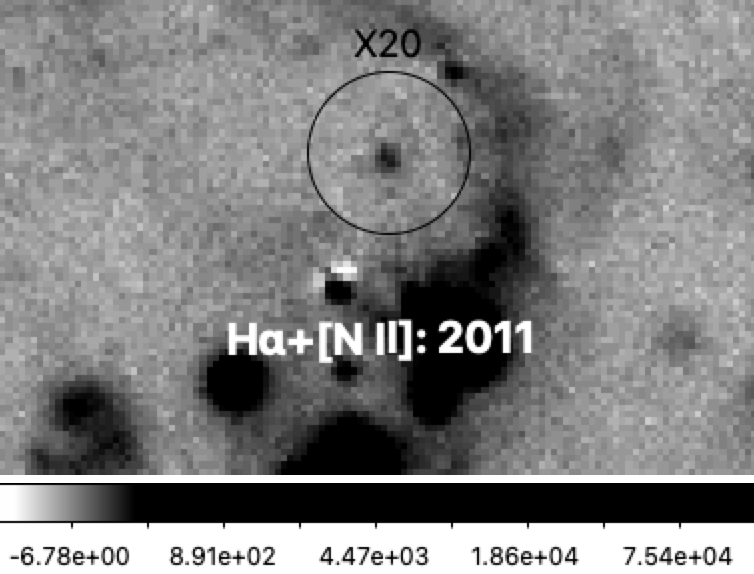}
  \caption{The source X20 is the Supernova 2008bk. In the left and middle panels we see the {\it{Chandra}} images from 2003 (OBSID: 3954) and 2011 (OBSID: 14231) respectively in the broad band (0.5 - 7 keV) and on the right the continuum subtracted H$\rm \alpha$+[\ion{N}{II}] image from Blanco 4m telescope (MOSAIC II camera)  obtained in 2011. In all figures the radius of the circle is 3.5 arcsec.}\label{fig:SN2008bk}
\endminipage
\end{figure*}

Source X20  coincides with the Hydrogen rich SN 2008bk and it appears in the {\textit{Chandra}} observations only after 2008. SN 2008bk is a low-luminosity SN II-P (e.g. \citealt{2012AJ....143...19V}). Its progenitor is a red supergiant with an initial mass of $\sim$ 13 M$\odot$ (\citealt{2014MNRAS.438.1577M}), while earlier a mass of  $\sim$ 8.5 M$\odot$ had been reported (\citealt{2008ApJ...688L..91M}; \citealt{2012AJ....143...19V}). In \autoref{fig:SN2008bk} we show on the left observations of that region in 2003, and in the middle observations in 2011 with {\textit{Chandra}}, and on the right an inverted-color H$\rm \alpha$+[\ion{N}{II}] image from Blanco 4m telescope (MOSAIC II camera)  obtained in 2011. In the latter we can also see  the light echo around it \citep{2013AJ....146...24V}. We note that at this phase, we observe the SN 2008bk in both {\textit{Chandra}} and optical data at an offset of 1.7 arcseconds compared to the position reported by \citet{2008ATel.1448....1L}. Hence, we refine its position and the new coordinates are RA: $23h57m50.5s$ and  DEC: $-32d 33m 20.3s$.

The interaction between the SN ejecta and CSM can produce X-ray emission, as observed in the hydrogen-rich SN 1979C (e.g. \citealt{1982ApJ...259..302C}). Recent radiative transfer simulations of hydrogen-rich SNe (Type II SNe) further illustrate how interaction-driven power influences their optical spectra at late phases (>1000 days). These simulations suggest that interaction primarily enhances flux in the ultraviolet and, if the shock power is not fully thermalised, the SN should exhibit significant X-ray luminosity \citep{2023A&A...675A..33D}. 

Although SN 2008bk appears very faint in the {\it{Chandra}} observations, we detect it in all the observations of 2011, i.e. in the OBSIDs: 14231, 13439, and 14378, with the highest S/N (in the broad band) in this of 14231 where S/N = 2.76. We do not detect the SN 2008bk in later observations, most probably because of the significantly less exposure time. Because of the faintness of the supernova, we could not extract any spectrum using {\it{Chandra}} observations. Hence, we used PIMMS in order to estimate the flux. To do that we have considered i) a thermal emission model with $kT=1.5\,keV$ (e.g., \citealt{2002ApJ...572..932P}) and ii) a power law with an index of 1.9 (e.g., \citealt{2007MNRAS.381..280M}). The fluxes are $(1.21\pm0.32)\times10^{-15}\, erg\,s^{-1}\, \,cm^{-2}$ and $(2.37\pm0.63)\times10^{-15}\, erg\,s^{-1}\, \,cm^{-2}$ for the two models respectively.

\subsubsection{X-ray sources known or suspected as Supernova Remnants}
In this subsection we compare the detected X-ray sources with known or suspected as SNRs from other studies. Sources X11, X13, X15, X25, and X38 coincide with optical SNRs and hence, this analysis focuses on them. From those, only X15 has been detected before and is a known X-ray SNR. 

\begin{figure}
\minipage{0.5\textwidth}
  \includegraphics[width=0.48\linewidth]{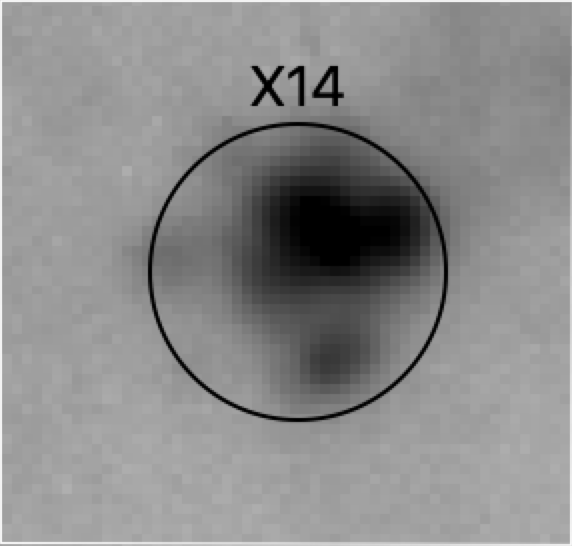}
\hfil
 \includegraphics[width=0.48\linewidth]{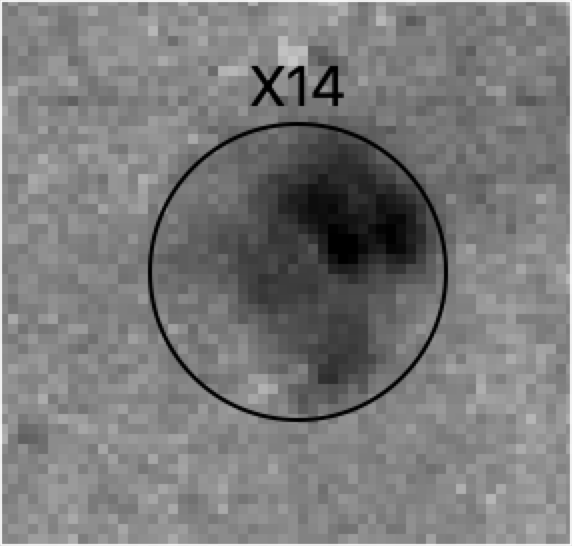}
  \caption{The optical counterpart of the source X14 in H$\rm \alpha$ + [\ion{N}{II}] (left) and [\ion{S}{II}] (right). The optical images were obtained from Blanco 4m telescope (MOSAIC II camera) in 2011.} \label{fig:X14}
\endminipage
\end{figure}

Source X14 (CXOU J235746.7-323607 in \citealt{2011AJ....142...20P}) is the X-ray counterpart to the candidate radio SNR R3 \citealt{2002ApJ...565..966P}). This source was presented earlier as P10 by \citet{1999A&A...341....8R} who suggest that its soft, X-ray spectrum come from a super-bubble or multiple SNRs. However, P10 has an offset of 6'' with source X14, hence they are probably two different sources, physically associated. \citet{2011AJ....142...20P} reported that its time variability rejects the single-SNR scenario and it may be a SNR/XRB system analogous to the Galactic source W50/SS 433 (\citealt{1999ApJ...512..784S}), where synchrotron emission results from the collision between the jets from the microquasar SS 433 and the surrounding SNR (W50) shell.  In \autoref{fig:X14} this source in H$\rm \alpha$+[\ion{N}{II}] and [\ion{S}{II}] from MOSAIC II camera in Blanco 4 meter telescope (Chile) are presented. The [\ion{S}{II}]/H$\rm \alpha$ ratio is $\sim$ 0.15 (calculated using the photometry presented in \citealt{2021MNRAS.507.6020K}) and it does not satisfy the traditional diagnostic for being a SNR, according to which,  this ratio is higher than 0.4 for SNRs. In order to explore if this is indeed an optical SNR,  other diagnostics that require more emission lines or spectra are needed. (e.g \citealt{2020MNRAS.491..889K}). However, as can be seen in \autoref{fig:X14}, there are multiple "peaks", perhaps indicating that there are more sources there. A contaminating source, bright in H$\rm \alpha$, would lead to lower [\ion{S}{II}]/H$\rm \alpha$ ratios, or other ratios indicative of shock excitation, even if there is an SNR there. Finally, it could be a young SNR, the shock velocity of which, and so its temperature, are too high to produce strong optical emission lines, such as [\ion{S}{II}] and [\ion{N}{II}]. In order to be conservative, we do not consider X14 as SNR.

 
Source X9 coincides with the source CXOU J235743.8-323633 in \citet{2011AJ....142...20P}. It has an offset of 2.5'' from the H II region D22, reported in \citet{1980ApJ...242...30D} (D22). \citet{2011AJ....142...20P} suggested that it can be either a SNR or a X-ray binary associated with the H II region. X9 does not present any emission in the optical, and it is close to the edge of a large, ring-like structure which has not been identified as SNR. Hence, at this point, we do not classify X9 as an SNR.

X15 has been identified as an optical SNR in \citeauthor{Blair1997} (\citeyear{Blair1997}; S11),  in \citeauthor{2021MNRAS.507.6020K} (\citeyear{2021MNRAS.507.6020K}; 7793\_22), and in \citeauthor{2024MNRAS.530.1078K} (\citeyear{2024MNRAS.530.1078K}; NGC7793\_SNR\_136), as a radio SNR in \citeauthor{2002ApJ...565..966P} (\citeyear{2002ApJ...565..966P}; N7793-S11), and X-ray SNR  in \citeauthor{2011AJ....142...20P} (\citeyear{2011AJ....142...20P}; CXOU J235747.2-323523). Sources X11, and X13, coincide with the optical SNRs 7793\_24 and 7793\_23 in \citet{2021MNRAS.507.6020K} and NGC7793\_SNR\_91 and NGC7793\_SNR\_126 in \citet{2024MNRAS.530.1078K}. Source X25 (J235752.2-323413 in \citealt{2011AJ....142...20P}) coincides with the optical SNR 7793\_5 in \citet{2021MNRAS.507.6020K}, and X38 with the S24 in \citet{Blair1997} and 7793\_21 in \citet{2021MNRAS.507.6020K}. In the study of \citet{2012MNRAS.419.2095M} the sources X11 and X25 are presented as J235743.9-323441 and J235752.2-323413 respectively, and they are reported to reside in areas dominated by high-mass X-ray binary (HMXB) population. The offset of X11, X13, X15, X25, and X38 from the optical SNRs are: 0.7, 1.3, 1.3, 0.4, and 0.6 arcseconds respectively. The coordinates of the optical SNRs have been taken from \citeauthor{2024MNRAS.530.1078K} (\citeyear{2024MNRAS.530.1078K}; X11, X13, X15) and \citeauthor{2021MNRAS.507.6020K} (\citeyear{2021MNRAS.507.6020K}; X25, X38). 

\subsection{X-ray colors}
We further examine the properties of X11, X13, X15, X25, and X38, that coincide with optical SNRs, in order to decide if they should be indeed considered as X-ray SNRs. The most accurate way to do so, is to examine their X-ray spectra. X-ray SNRs present thermal emission from shock-heated plasma, and sometimes, young SNRs non-thermal emission from relativistic electrons  moving in the magnetic fields of hot plasma. The thermal emission is often characterized by emission lines of  O, Ne, Mg, Si, S, Ar, Ca and iron-group elements (e.g \citealt{2020pesr.book.....V}). However, the faintness of SNRs at distances such as NGC 7793, usually does not allow the extraction of a detailed spectrum where emission lines can be observed, and hence no emission lines are present. 

\begin{figure*}
\centering
\includegraphics[width=\textwidth]{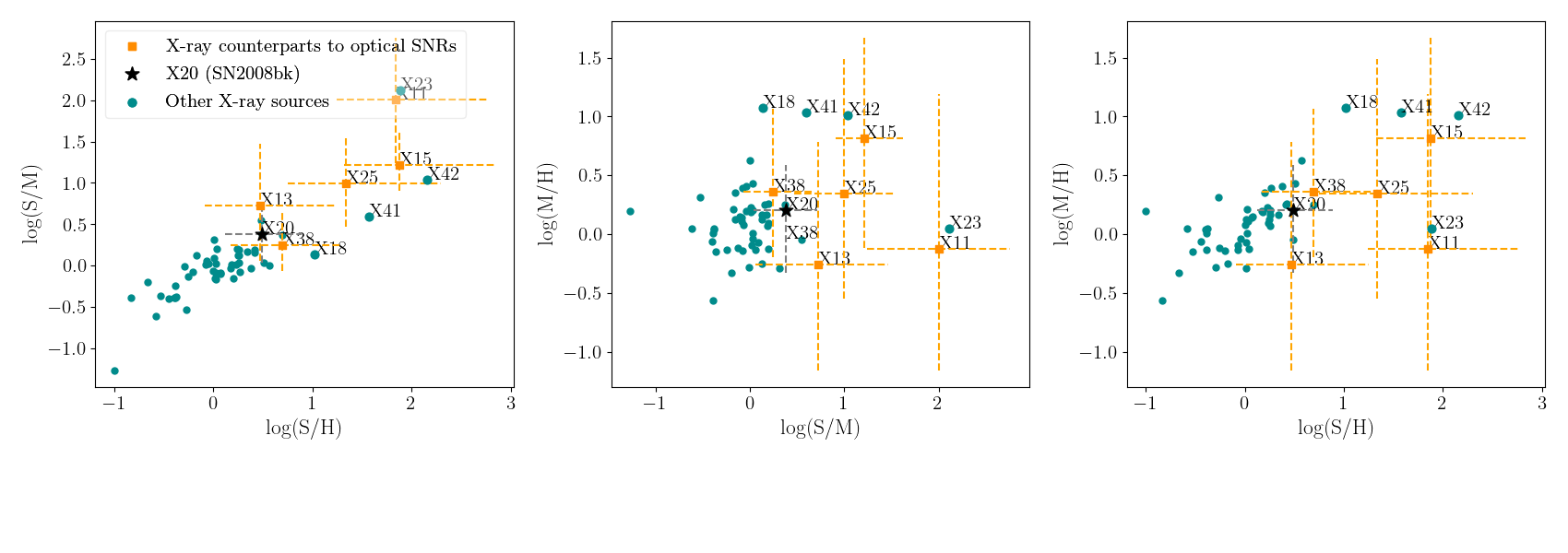}
\caption{\label{fig:hr_all_grids} The $\rm log_{10}(S/M)-log_{10}(S/H)$, $\rm  log_{10}(M/H)-log_{10}(S/M)$, and $\rm  log_{10}(M/H)-log_{10}(S/H)$ X-ray colors for all the X-ray detected sources with S/N > 1, detected in this work. The orange squares correspond to the X-ray sources that coincide with optical SNRs. }
\end{figure*}

In such cases, colors or hardness ratios are a good proxy to quantitatively characterize a spectrum. In this study, We calculated the colors $\rm log(S/M),  log(S/H) ,\, and\ log(M/H)  $ as described in \S \ref{sec:HR}. Then the color-color diagrams $\rm log(S/M) - log(S/H)$, $\rm log(S/M) - log(M/H)$, and $\rm log(S/H) - log(M/H)$ were constructed and are presented in \autoref{fig:hr_all_grids}. All the data points refer to the X-ray sources with S/N > 1 in the soft band. Orange squares are the X-ray sources X11, X13, X15, X25, and X38,  that coincide with optical SNRs, while the star is the SN2008bk (X20). As can be seen, although the error-bars are large, in all three plots, the orange squares occupy the soft part of the diagrams. This is the region where X-ray SNRs are expected since they are characterized by temperatures up to $\rm \sim 10^7\, K$.

\subsection{New X-ray supernova remnants}
In this study, we detected X-ray sources in the galaxy NGC 7793 and we found that 5 of them (X11, X13, X15, X25, X38) coincide with optical SNRs (one is a known X-ray/radio SNR) with offsets smaller than 1.3 arcsec.   The X-ray colors of these sources present soft or super-soft emission, which is expected from SNRs. 

We examine the probability of these sources to present long or short variability. For the long variability we compared the photometry of the different observations where these sources have been detected and we found that it is stable within the uncertainties. For the short-variability check the tool \texttt{glvary} of the \texttt{CIAO} software was used. This tool looks for significant deviations among multiple time bins of the events, using the Gregory-Loredo algorithm. Among the output values,  the probability of the examined source to be variable is given. If the variability index that returns the \texttt{glvary} is less than 3, we can consider our sources as non-variable. All the 5 sources we  examine gave variability index less or equal to 2. More specifically, those are 1, 0, 0, 1, and 2, for X11, X13, X15, X25, and X38 respectively. No variability is expected from SNRs, so the lack of variability observed in these sources enhances the SNR scenario. 

Concluding, X11, X13, X25, X38 can be considered as new, X-ray SNRs, additional to the known SNR X15. For consistency and comparison reasons, in the following analysis we have included the SNR X15, although it is a known X-ray SNR (\citealt{2011AJ....142...20P}). These X-ray SNRs had not been presented before because: (a) they were not detected in other X-ray studies. In this study we ran the detection on individual but also combined observations, increasing the S/N; and/or (b) no optical/radio identification had been reported before (the absence of which makes very uncertain its identification as an X-ray SNR).

\subsection{Candidate X-ray SNRs without optical counterpart}
In \autoref{fig:hr_all_grids}, apart from X11, X13, X15, X25, and X38, there are more sources that fall in the soft part of the color-color diagram. Those are X18, X23, X41 and X42. We examine these sources in order to decide if they could be considered as  candidate SNRs. X18 is the foreground star P6 (\citealt{1999A&A...341....8R}, \citealt{2003AJ....125..984M}, \citealt{2011AJ....142...20P}) that earlier had been misclassified as an H II region (H18; \citealt{1969ApJS...18...73H}). X41 is a microquasar, as discussed also in section \S \ref{sec:comparison_ap} 
 (\citealt{2010Natur.466..209P}, \citealt{2010MNRAS.409..541S}). Around the microquasar there is a super-bubble (e.g \citealt{2021MNRAS.507.6020K}) the emission properties of which are similar to a SNR, that is why first it had been misclassified as a SNR (S26 in \citealt{Blair1997}).

X23 (detected for the first time in this study) and X42 have not been previously classified but have been reported near regions dominated by X-ray binaries (\citealt{2012MNRAS.419.2095M}). They appear in the soft part of the color-color diagrams (\autoref{fig:hr_all_grids}), and they do not present any variability. Putting everything together, we do not have any strong evidence against the scenario of X23 and X42  to be SNRs, and hence we can consider them as candidate SNRs. The reason that we do not detect any optical emission can be an implication of a low density environment, where no bright optical emission is produced. This in turn could imply a SN Ia origin (e.g., \citealt{2021ApJ...908...80W}).

\subsection{X-ray spectral analysis of SNRs and candidate SNRs }
Sources that are described in the previous sections as known or candidate SNRs (X11, X13, X15, X23, X25, X38, X42) are very soft and have very few counts in the \chandra{} data, not allowing for spectral modeling. 
While the \chandra{} X-ray colors offer valuable insights into the spectral properties of the sources, we aimed to take this a step further by extracting spectra from archival \xmm{} observations. This approach allows us to take full advantage of \chandra{}'s high spatial resolution for precise source positions, while leveraging \xmm{}'s large effective area and better sensitivity at soft energies to maximize the spectral information obtained from these sources.
To put this into perspective, the brightest of these sources, X15, a known supernova remnant, has approximately 99 combined net counts with \chandra{}, while the combined EPIC MOS net counts, exceed this by more than six times, reaching around 600, allowing for detailed spectral modeling with \chisq square statistics.

We used 14 \xmm{} observations (listed in \autoref{table:xmm_info}) that covered the sources without contamination from any nearby outbursting sources. For the standard calibration of the observations we utilized the \xmm{} Science analysis system (SAS) v19.0.0 and obtained filtered event files clean from background flares as described in detail in \citet{haberl12} and \citet{konna23}. 
During this study we are using only the EPIC MOS detectors which offer a combined clean exposure time of $\sim$1.1Msec. Although the PN detector offers higher sensitivity, we found that the combined MOS spectrum provided comparable S/N and lower background, particularly at soft energies ($<
$1\,keV), allowing clearer detection of the spectral emission lines.

\begin{table}[h!]
	\centering
		\caption{\xmm{} observations of NGC\,7793 utilized in this study} 
	\begin{tabular}{lccc} 
		\hline
		OBS ID & Exp MOS1  & Exp MOS2 & Start date\\
		       & sec & sec &   \\
               \hline
0693760401 & 47617  & 47585& 2013-11-25  \\
0748390901 & 48649  & 48620& 2014-12-10  \\
0804670301 & 54885  & 55063& 2017-05-20  \\
0804670401 & 32170  & 32513& 2017-05-31  \\
0804670501 & 32910  & 32991& 2017-06-12  \\
0804670601 & 31095  & 31117& 2017-06-20  \\
0804670701 & 51643  & 51618& 2017-11-25  \\
0823410301 & 25938  & 26078& 2018-11-27  \\
0823410401 & 26587  & 26572& 2018-12-27  \\
0853981001 & 41669  & 42373& 2019-05-16  \\
0840990101 & 38269  & 38716& 2019-11-22  \\
0861600101 & 67413  & 67666& 2020-06-27  \\
0883780101 & 40524  & 40573& 2021-05-29  \\
0883780201 & 30476  & 33721& 2021-11-20  \\
		\hline
	\end{tabular}
    \tablefoot{The exposure times correspond to clean exposures after the removal of time periods with high background flares.}
	\label{table:xmm_info}
    \end{table}

We extracted the source spectra for each OBSID MOS detector using the \textit{evselect} SAS task, centering the extraction regions on the \chandra{} source coordinates. The extraction regions were carefully chosen to enclose at least 80\% of the PSF while minimizing contamination from nearby sources. The background regions were placed in source-free areas near the targets and were sufficiently large to ensure an adequate number of counts in each observation. We then extracted the Auxiliary Response Files (ARF) for the source spectra and the Redistribution Matrix Files (RMF) using the SAS task \textit{arfgen} and \textit{rmfgen} respectively. The background area was calculated using the \textit{backscale} SAS task. 

To maximize the signal-to-noise ratio, we combined all MOS spectra using the SAS task \textit{epicspeccombine}, which generates a combined source spectrum, background spectrum, and response file. To fit the spectra we used the  XSPEC v.12.13.0c software \citep{arnaud96} an two models typically used to fit the shock heated plasma that produces the X-ray emission in SNR spectra, namely the \texttt{vapec} and \texttt{vpshock} models. The model \texttt{vapec}  corresponds to collisionally-ionized diffuse gas in thermal equilibrium  while the \texttt{vpshock} model is a constant temperature plane-parallel shock plasma model that allows for non-equilibrium.
Both models provide the possibility to vary the abundances.
In order to allow for \chisq statistics during the X-ray spectral fitting process we grouped the data in bin of at least 20 total counts. 
We used a model of the form \texttt{tbabs$\times$phabs$\times \sum$\texttt{model}} where \texttt{model} corresponds to either the \texttt{vapec} or the \texttt{vpshock} model. 
The \texttt{tbabs} component was used to model interstellar medium (ISM) absorption, with its value fixed at $\rm 3.4\times10^{20} \,atoms\, cm^{-2}$ $-$ the weighted average absorption toward NGC\,7793. The second absorption component, \texttt{phabs}, was allowed to vary to account for local ISM absorption fluctuations or absorption intrinsic to the supernova remnant (SNR). In all cases, we adopted the abundance table from \citet{wilms00}, as it is the most up-to-date reference for use with the \texttt{tbabs} model.

\begin{figure}[!htbp]
 	\centering
\includegraphics[width=1.0\columnwidth]{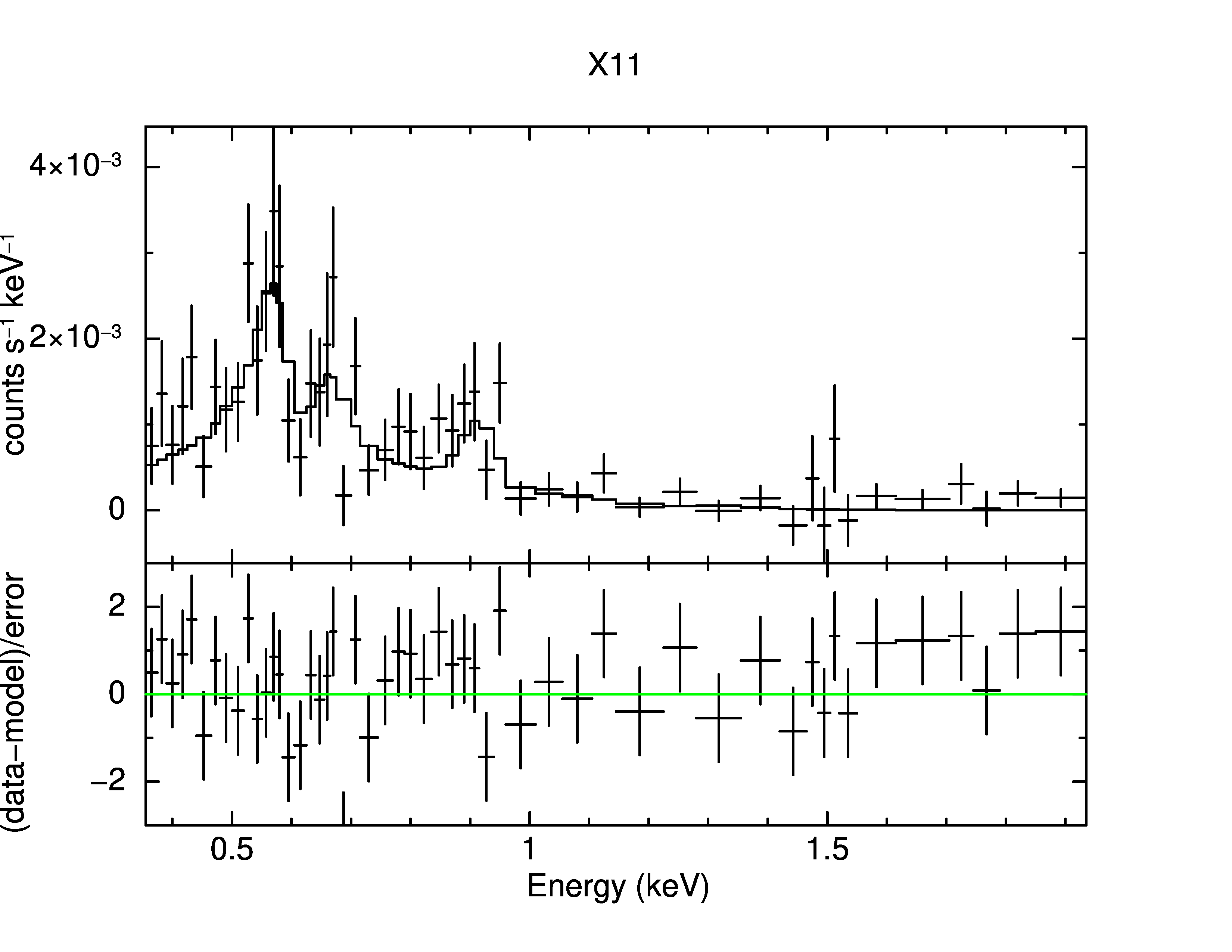}  
\includegraphics[width=1.0\columnwidth]{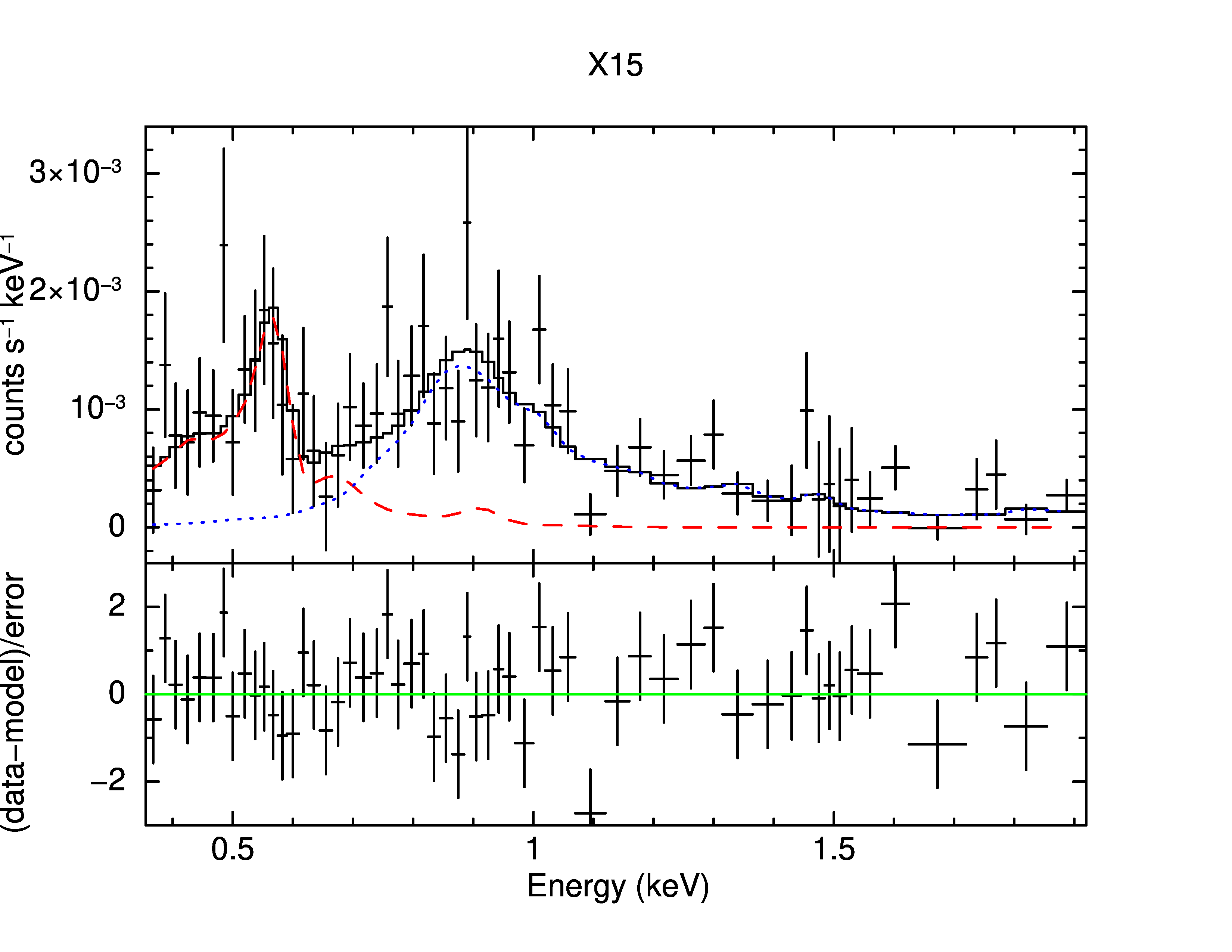}
\includegraphics[width=1.0\columnwidth]{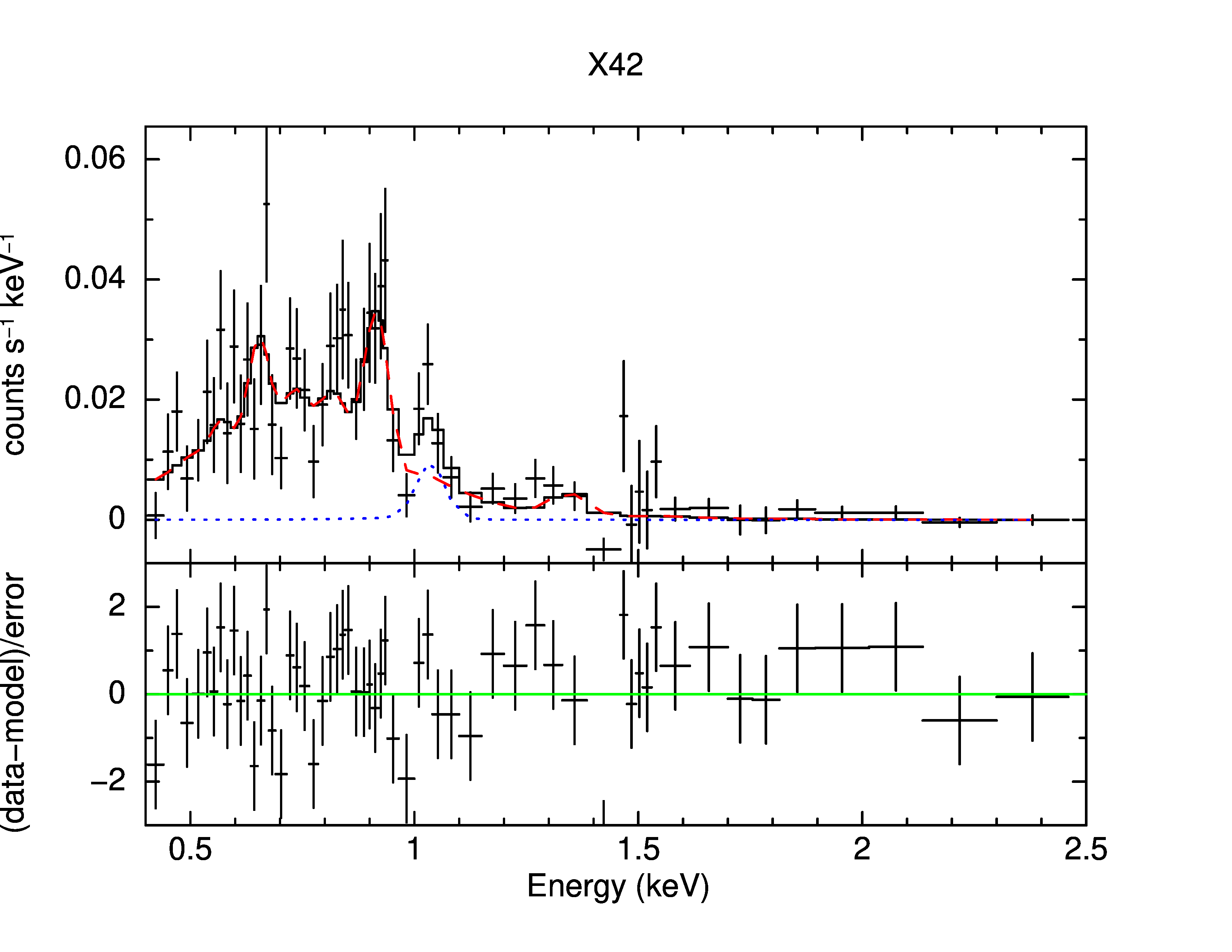}
           \caption{Combined EPIC MOS spectra and best-fit models for sources X11, X15 and X42.
           The spectra of X11 and X42 are fitted with one temperature thermal plasma model. The spectrum of X15 is fitted with two temperature thermal plasma components shown with red dashed and blue dotted lines. The fit residuals for all sources are displayed in the bottom panels of each plot, with error bars representing 1$\sigma$ uncertainties.}
 	\label{fig.spectraxmm}
 \end{figure}

We successfully extracted and modeled spectra for sources X11, X15, and X42. However, sources X25 and X38 were too faint for spectral modeling, while sources X13 and X23 were located near other X-ray sources, making it impossible to disentangle their emission.
In \autoref{tab.xmmfits} we present for sources X11, X15, and X42, only the results for the \texttt{vapec} model since the \texttt{vpshock} model yielded results consistent with a plasma near equilibrium. For source X42, we thawed the Ne and Mg abundances and included an additional Gaussian component (\texttt{gauss}) at 1.03\,keV to better account for the emission lines present in the spectrum. We present the best-fit parameters along with the corresponding fluxes and luminosities, with uncertainties given at the 1$\sigma$ level. We present the combined \xmm{} MOS spectra along with their best-fit models in \autoref{fig.spectraxmm}.

All sources are soft emitters with no detected emission above 2 keV, and their absorption components agree within the uncertainties, suggesting similar conditions at their locations.
The spectra exhibit strong emission lines from K-shell transitions of various elements, indicating a thermal plasma with  temperature exceeding $2.5 \times 10^6$ K. Specifically, in X11, we detect prominent emission lines of O~VII ($\sim0.55$ keV), O~VIII ($\sim0.65$ keV), and Ne~IX ($\sim0.9$ keV). In X15, we observe O~VII ($\sim0.55$ keV) and Ne~IX ($\sim0.9$ keV), while in X42, O~VIII ($\sim0.65$ and $0.85$ keV) and Ne~IX ($\sim1.05$ keV) are present.

Source X11 is well fitted with a single soft thermal component at $kT = 0.13$ keV. In contrast, a single thermal component did not provide a good fit for source X15, yielding $\chi^2_{\nu} \approx 2.0$. Instead, the best-fit model required two thermal plasma components: a soft component with $kT < 0.12$ keV and a harder component at $kT = 0.78$ keV. Source X42 is well described by a single thermal component at $kT = 0.19$ keV, though residuals remain around $\sim1.5$ keV (Ne~IX).

Over the 8 years of available \xmm{} observations, we detect no long-term variability, which is consistent with expectations for SNRs.

The low temperatures ($kT<0.78\, keV$) of the best-fit spectra (\autoref{tab.xmmfits}) suggest that these are relatively old SNRs. This is also supported by the fact that they exhibit strong optical emission.
\begin{table*}[!h]
	\centering
		\caption{\xmm{} spectral modeling}
        \small
            \begin{tabular}{@{}lllllcccc@{}}
			\toprule		
		Name &  $\rm{N_{H}^{local}}$ & kT & norm   &$\chi^2_{\nu}$ ($\chi^2$/dof)  & F$_X$ &F$\mathrm{_X^{ISMcor}}$(F$\mathrm{_X^{cor}}$) &  L$_X$ &L$\mathrm{_X^{ISMcor}}$(L$\mathrm{_X^{cor}}$) \\[2pt]
  &  $10^{22}$  \cmsq  &  keV  &$10^{-4}$ && $10^{-15}$\funit & $10^{-15}$\funit &$10^{36}$\ergs &$10^{36}$\ergs\\
 (1)  &  (2)  &  (3) & (4)& (5) & (6) & (7) & (8) & (9)\\
\midrule
\multicolumn{9}{c}{XSPEC model: \textit{tbabs$\times phabs \times \sum$vapec} } \\[3pt]
X11  & 0.48$_{-0.30}^{+0.35}$ & 0.13$\pm 0.04$ &1.5$_{-1.5}^{+102.0}$  &1.19 (58.69/49)&3.35$_{-3.35}^{+0.01}$&3.96 (5.05) &5.49$_{-5.49}^{+0.01}$ &6.49 (8.27) \\ [4pt]
X15  & 0.46$_{-0.24}^{+0.34}$ & $<0.12$ &8.5$_{-8.1}^{+58}$ & 0.95 (50.51/53)&3.92$\pm1.8$ &4.38(32.7) &6.42$\pm2.94$ &7.17 (53.6) \\ [4pt]
  &  & 0.78$_{-0.13}^{+0.09}$&0.03$_{-0.01}^{+0.02}$ & & \\ [4pt]
X42\tablefootmark{a}  & 0.21$_{-0.19}^{+0.30}$& 0.23$\pm 0.04$ & 0.09$_{-0.05}^{+0.42}$  &1.32 (68.89/52)&3.87$_{-3.60}^{+0.01}$ & 4.36 (9.71) &6.34$_{-5.89}^{+0.41}$ & 7.14(15.8) \\ [4pt]

\bottomrule
    \end{tabular}
    \tablefoot{
    In Column 1, we provide the name of the source spectrum. Columns 2 to 4 list the best-fit component values: the local absorption, the temperature of the thermal plasma, and its normalization, respectively. The Galactic column density towards NGC\,7793 is fixed to a value of $3.4 \times 10^{20}~\text{atoms}~\text{cm}^{-2}$. The normalization of the thermal component is expressed in units of 
$ \frac{10^{-14}}{4\pi D^2} \int n_e n_H \, dV $, 
where $n_e$ and $n_H$ are the electron and hydrogen densities, respectively, integrated over the emitting region's volume $V$, and $D$ represents the source distance in centimeters \citep{smith01}. 
In Column 5, we present the values of the $\chi^2$ statistic. Column 6 and 7 provide the absorbed 0.5--7.0\,keV flux, and the corrected values (for only the ISM and for total absorption) respectively. Columns 8 and 9 contain the corresponding luminosity values. Errors are quoted at the $1\sigma$ confidence level. The adopted distance to NGC\,7793 is $3.7~\text{Mpc}$. \tablefoottext{a} The abundances of the following elements were left free to vary Ne=$2.21^{+1.27}_{-0.88}$, 
Mg=$2.94^{+4.08}_{-2.78}$, and a Gaussian line was added to the model with best fit parameters LineE=$1.03\pm 0.02$~keV and 
norm = $(1.66^{+1.29}_{-1.11}) \times 10^{-7}$. }
    \label{tab.xmmfits}
\end{table*}

\subsection{Exploring correlations with optical properties} \label{sec:correlations}

X-ray SNRs are usually young SNRs, where shock velocities are
very high and heat the ISM to temperatures of $\rm 10^7\,K $ producing thermal X-ray emission. On the other hand, optical emission, especially from forbidden lines, such as [\ion{S}{II}] and [\ion{N}{II}] is expected mainly from more evolved SNRs, where the temperature has dropped to $\rm \sim 10^4\, K$. The presence of both thermal X-ray and optical emission is probably an indication of a non-uniform or very dense environment. For example, a highly non-uniform environment may permit a part of the SNR to expand with high velocity, emitting in X-rays, while another part of the shock, evolving in a denser environment, to have already entered the radiation phase. Another scenario for having both optical and X-ray emission in SNRs can be the case of mixed morphology SNRs, where optical/radio emission is observed from the shell, while thermal, X-ray emission from the center. This kind of SNRs require a dense environment (e.g \citealt{2024MNRAS.531.5109C}). The lack of the spatial resolution in the study of galaxies out of the Local Group, prevents us from examining the aforementioned scenarios. 

Subsequently, we examine possible correlations between X-ray and optical properties of SNRs.  In order to account for the instrument's characteristics (e.g., effective area), we choose to perform the comparisons using X-ray luminosities rather than raw count rates. However, spectral fits are available for only 2 out of the 5 SNRs (\autoref{fig.spectraxmm}), for which the $kT$ and $N_H$ parameters have been determined. For the remaining SNRs, we assumed a $kT = 0.5\,keV$, a typical value for a thermal  bremsstrahlung model for SNRs (e.g. \citealt{2010ApJ...725..842L}), and a $N_H = 0.504\times10^{22}\,cm^{-2}$, the average of the total $N_H$ values for X11 and X15 in \autoref{tab.xmmfits} (Galactic + local). We chose only X11 and X15 (and not X42) because they exhibit optical emission, and we expect similar properties for the also optically detected X13, X25, and X38. Given that unabsorbed fluxes are highly model-dependent, and small deviations from the true values of $kT$ and $N_H$ can result in significantly different estimates, we use and present the observed fluxes, rather than those corrected for local and Galactic extinction. The fluxes were calculated using the PIMMS tool (detailed description in \S \ref{sec:detection}). The observed luminosities, along with other relevant properties, are listed in \autoref{tab:x_opt_prop}. For the calculation of the luminosity, and throughout this study, we adopted a distance of 3.7 Mpc \citep{2011ApJS..195...18R}.

We explore the relation between X-ray emission and colors with the H$\rm \alpha$ luminosity, the emission line ratios [\ion{S}{II}]/H$\rm \alpha$, [\ion{N}{II}]/H$\rm \alpha$, and [\ion{O}{III}]/H$\rm \beta$ (these ratios are indicative of the shock excitation of the gas), and the densities that have been calculated using the ratio [\ion{S}{II}](6717)/[\ion{S}{II}](6731). The main idea behind the use of the [\ion{S}{II}] ratio for the density estimation, is that the collisional excitation from lines from the same ions,  with  more or less the same excitation energy, 
is proportional to the collision rates, and hence their ratio is a density indicator (e.g \citealt{2006agna.book.....O}; \citealt{2011piim.book.....D}).  Due to the lack of an optical spectrum of X25, the density has not been calculated. In \autoref{tab:x_opt_prop} all these properties are presented. The H$\rm \alpha$ flux and the optical emission-line ratios of X11, X13, X15 have been obtain from \citet{2024MNRAS.530.1078K}, for  X25 from \citet{2021MNRAS.507.6020K}, and the X38 from \citet{Blair1997} where they present spectroscopic fluxes and ratios. We note here that for the latter, the [\ion{O}{III}] and [\ion{N}{II}] emission includes emission from the 4959\AA\ + 5007\AA\ and 6548\AA\ + 6584\AA\ lines respectively, while for the rest of the cases the emission comes from the [\ion{O}{III}](5007) and [\ion{N}{II}](6583). 

\begin{table*}[htbp]
\centering
\caption{Optical and \chandra{} X-ray properties of the X-ray SNRs}
\begin{tabular}{llllllll}
\hline
ID  & L$_{X}$ & $\rm log_{10}(S/H)$ & L$\rm _{H\alpha}$  & [\ion{S}{II}]/H$\rm \alpha$ & [\ion{N}{II}]/H$\rm \alpha$  & [\ion{O}{III}]/H$\rm \beta$ & Density (d) \\
& ($\times 10^{36}\,erg\,s^{-1}$) & & ($\times 10^{36}\,erg\,s^{-1}$) & & & & ($cm^{-3}$) \\
\hline
X11 &  5.49$^{^{+0.01}_{-5.49}}$  & 1.84$^{+0.92}_{-0.6}$           &15.0$\pm$0.05    &0.71     &0.46  &7.08   &620.96 \\
X13 &  0.88$\pm$0.45  & 0.47$^{+0.78}_{-0.56}$          &11.0$\pm$0.05  &1.07     &0.5   &2.45   &6.72  \\ 
X15 &  6.42$\pm$2.94  & 1.88$^{+0.96}_{-0.56}$          &6.6$\pm$0.03   &1.15     &0.6   &1.82   &483.97  \\
X25 &  1.79$\pm$0.26  & 1.34$^{+0.96}_{-0.59}$          &1.6$\pm$0.05    &0.66     &-     &-      &-     \\
X38 &  1.49$\pm$0.47  & 0.69$^{+0.58}_{-0.52}$          &23.0$\pm$0.4   &0.61     &0.43  &2.97    &196.11 \\
\hline
\end{tabular}
\label{tab:x_opt_prop}
\end{table*}

\begin{figure*}
\minipage{\textwidth}
  \includegraphics[width=0.33\linewidth]{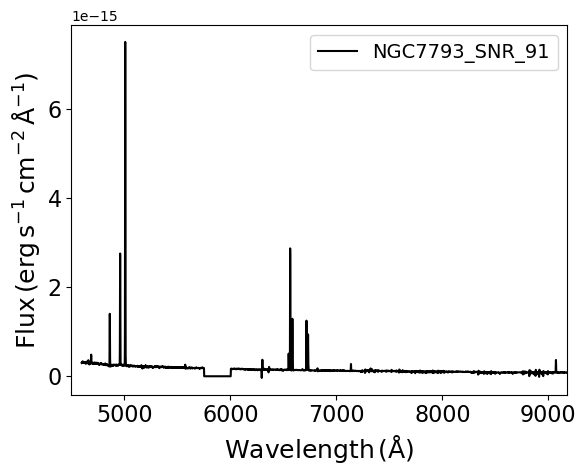}
\hfil
  \includegraphics[width=0.33\linewidth]{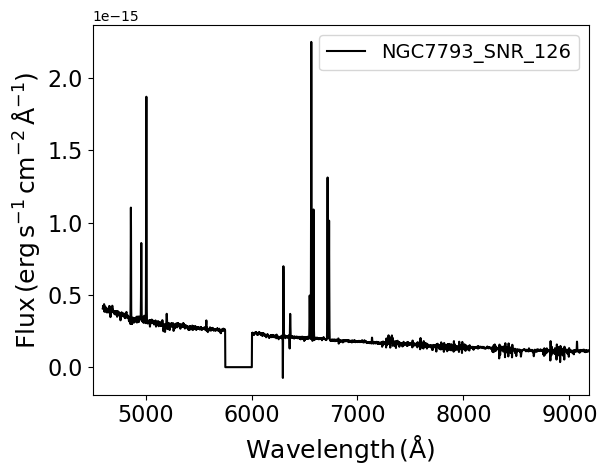}
\hfil
  \includegraphics[width=0.33\linewidth]{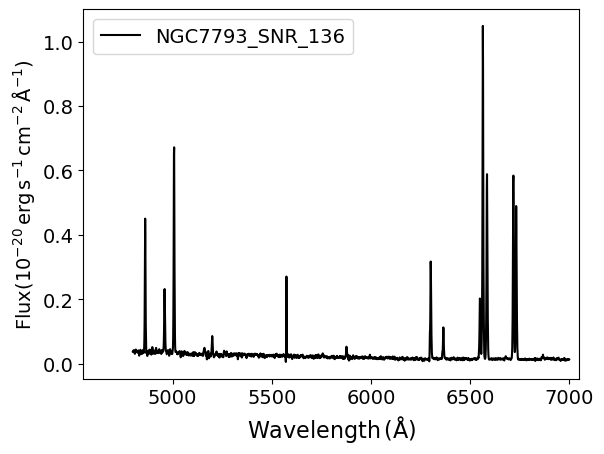}
  \caption{The optical spectra of the NGC7793\_SNR\_91, NGC7793\_SNR\_126, NGC7793\_SNR\_136, optical SNRs that spatially coincide with the X-ray sources X11, X13, X15 of this study.}\label{fig:opt_spec}
\endminipage
\end{figure*}

X11, X13, and X15 are the X-ray counterparts of NGC7793\_SNR\_91, NGC7793\_SNR\_126, and NGC7793\_SNR\_136, from the study of \citet{2024MNRAS.530.1078K}, the optical spectra of which are presented in \autoref{fig:opt_spec}. Spectral fluxes of X38 are also available from the work of \citet{Blair1997}. A common characteristic that we observe in the four of them is the strong [\ion{O}{III}] emission. Not all the SNRs present [\ion{O}{III}] emission. For example in the work of \citet{2024MNRAS.530.1078K}, only the $\sim$ 35\% of the sample presents [\ion{O}{III}] emission. This is most probably an indication that the shock velocity of these SNRs cannot be very low (e.g \citealt{1979ApJS...39....1R}). However, no correlation seems to exist between [\ion{O}{III}]/H$\rm \beta$ and X-ray luminosity or color.

\begin{figure*}
\minipage{\textwidth}
  \includegraphics[width=0.48\linewidth]{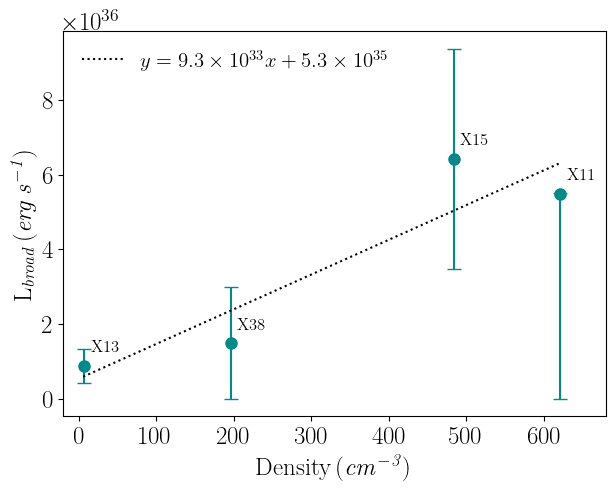}
\hfil
  \includegraphics[width=0.48\linewidth]{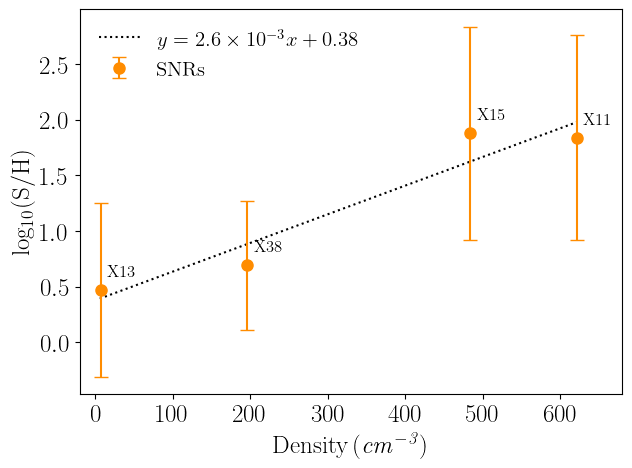}
\hfil
  \caption{\textit{Left:} The X-ray count rate in the broad band (0.5-7 keV), of the X-ray SNRs X11, X13, X15, and X38, as function of their density, calculated using the [\ion{S}{II}]6717/31 emission line ratio. \textit{Right:} The $\rm log_{10}(S/H)$ color of the same sources, as function of the density. In both cases the best-fit line is shown.}\label{fig:dens_plots}
\endminipage
\end{figure*}

Exploring all the parameters, we find a possible correlation between density and luminosity, and density and $\rm log_{10}(S/H)$, according to which luminosity and color increase with the density.  Both tendencies can be seen in \autoref{fig:dens_plots}. Although the sample is very small, only 4 points, we apply weighted Pearson correlation test (a method to measure a linear correlation between two variables, and weighted to take into account the uncertainties in luminosity and color), in order to examine if there is indeed any linear correlation and if this can be random or not. The results are presented in \autoref{tab:pearson}. Both relations show a correlation of around 0.9. However, the hypothesis of this correlation to be random cannot be excluded, as it is indicated by the relatively high p-values. More specifically, the probability of this association to be random is 34\% for the luminosity - density relation, and 14\% for the color-density relation. 

\begin{figure}
  \includegraphics[width=\linewidth]{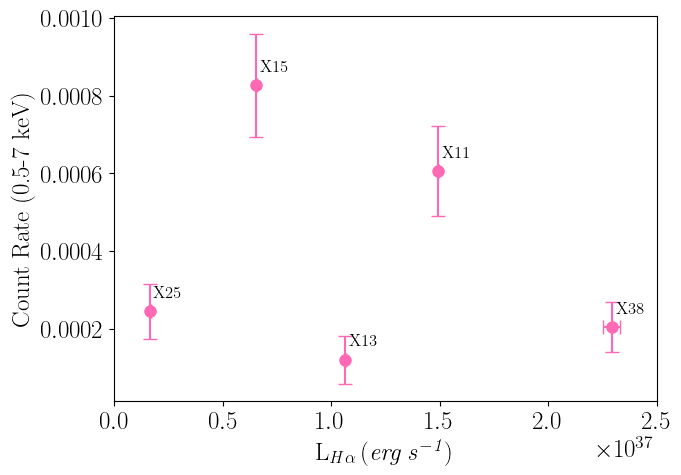}
  \caption{The  absorption corrected \chandra{} X-ray luminosity in the broad band (0.5-7 keV), of the X-ray SNRs suggested in this study, as a function of their H$\rm\alpha$ luminosity.}\label{fig:comp_fluxes}
\end{figure}

The increase of the X-ray luminosity with the density is theoretically expected. The X-ray and optical emission are related to the shock emissivity, which is proportional to the square of the density. If we had more sources, we could perhaps see a more quadratic relation than linear. In our case, the X-ray emission is dominated by the soft emission, and for this reason the color $\rm log_{10}(S/H)$ is always positive. Hence, similar behavior to the luminosity is expected for the relation $\rm log_{10}(S/H)$-density. Since both X-ray and optical emissions depend on the shock emissivity a correlation also between those two would be reasonable. However, such a correlation is not observed (\autoref{fig:comp_fluxes}). 
Absence of correlation between H$\rm \alpha$ and X-ray emission is also observed in the study of \citet{2007AJ....133.1361P}, where they compare the H$\rm \alpha$ luminosity of 9 optically identified SNRs in M101 and NGC 2403 with the X-ray luminosity of \textit{Chandra}-detected counterparts. The authors explain that this probably happens because of a non-uniform ambient that requires more complex emissivity models to be properly described. \citet{2013MNRAS.429..189L} reached a similar conclusion in their study of 16 optical and X-ray-emitting SNRs in NGC 4212, NGC 2403, and NGC 3077, where they also found no correlation. They support the idea that a non-uniform ISM, along with the coexistence of different materials at varying temperatures, likely contributes to the observed lack of correlation. In addition to a non-uniform ISM, the absence of correlation could also be influenced by the evolutionary phase of the SNRs. Younger SNRs are expected to exhibit stronger X-ray emission, whereas older SNRs tend to have stronger optical emission. To accurately examine these correlations, larger sample sizes are needed.

\begin{table}
\begin{threeparttable}
\caption{Pearson-Statistics parameters}
\begin{tabular}{ccccc}
\hline
\multicolumn{2}{c}{d - L$_{broad}$} & & \multicolumn{2}{c}{d - SH} \\ \cline{1-2}\cline{4-5}

r & p-value & & r. & p-value \\
\hline
0.85 & 0.40 & & 0.96 & 0.14 \\
\hline
    \end{tabular}
    \label{tab:pearson}
\begin{tablenotes}
\small{\item The first column refers to the correlation between density and X-ray luminosity in the broad band (0.5-7 KeV) and the second to the relation between density and the color $\rm log_{10}(S/H)$ where S is the soft band (0.5-1.2 KeV) and H the hard band (2 - 7 keV). In both cases r indicates the correlation coefficient.}
    \end{tablenotes}
\end{threeparttable}
\end{table}

\subsection{Comparison with X-ray SNRs in other nearby galaxies}

In this section, we compare our results with those of other galaxies. More specifically, in  \autoref{table:x_gal_info}, we list galaxies within 10 Mpc for which X-ray SNRs, identified as counterparts to optical SNRs, have been reported, along with their distance, SFR (based on the H$\rm \alpha$ luminosity of the galaxy), number of optical SNRs, number of X-ray SNRs, and the minimum X-ray luminosity of the X-ray SNRs. In general, more X-ray SNRs have been reported in these galaxies based on their soft emission and/or X-ray spectra, however, for consistency, we consider only those that have also been identified in the optical, and with the diagnostic [\ion{S}{II}]/H$\rm \alpha$ > 0.4. We apply this optical criterion for consistency, as optical SNRs in most of these galaxies have historically been selected based on this same criterion.

The purpose of this comparison is to estimate the expected number of X-ray SNRs in NGC 7793 and to identify the factors that may hinder their detection. We acknowledge that an objective comparison is not possible due to various biases. The most fundamental biases include variations in exposure times across surveys and differences in instrument sensitivities. In addition, there are cases where the X-ray data do not cover the entire galaxy. In addition, a low-density ISM/CSM, for example, resulting from low-metallicity stars with low mass-loss rates, could lead to non-detection.

\begin{figure}[!htbp]
  \includegraphics[width=\linewidth]{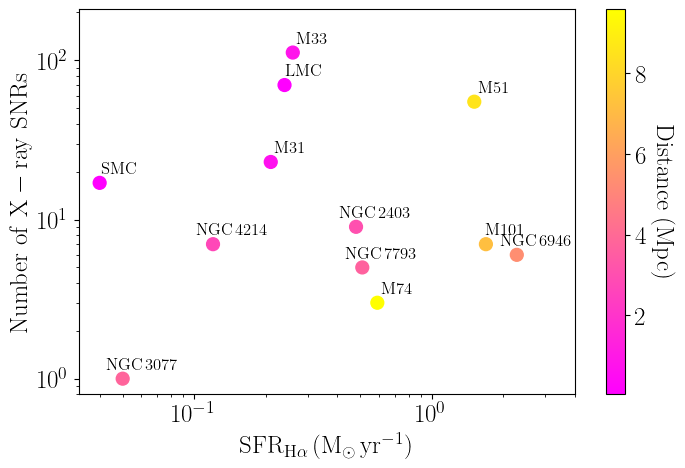}
  \caption{Number of X-ray SNRs, counterparts of optical SNRs with [\ion{S}{II}]/H$\rm \alpha$ > 0.4, as a function of their H$\rm \alpha$ star formation rate (SFR). The colorbar indicates the distance of the galaxies. }\label{fig:N_sfr}
\end{figure}

Examining the quantities presented in  \autoref{table:x_gal_info}, we do not find any strong correlations among them. In \autoref{fig:N_sfr}, we show the number of X-ray SNRs identified in the galaxy sample as a function of their SFR. The distance is shown in the colorbar. We observe a trend in which galaxies with higher SFRs, such as NGC 3077, NGC 4214, M31, LMC, and M33, tend to have a greater number of X-ray SNRs. These, along with SMC, are the nearest galaxies. However, M51, M101, M74, NGC 2403, NGC 7793, and NGC 6946 do not follow this trend. Surprisingly, M101, M51, and NGC 6946, which have significantly higher SFRs than the rest of the galaxies, present a relatively low number of X-ray SNRs. This is probably an indication that distance is a significant barrier to identifying X-ray SNRs, although we cannot make quantitative conclusions due to the biases mentioned earlier.

\begin{figure}[!htbp]
  \includegraphics[width=\linewidth]{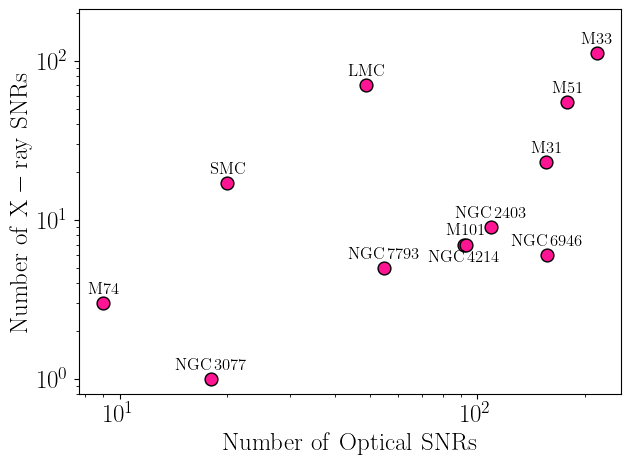}
  \caption{Number of X-ray SNRs, counterparts of optical SNRs with [\ion{S}{II}]/H$\rm \alpha$ > 0.4, versus the number of optical SNRs }\label{fig:N_opt_xrays}
\end{figure}

In \autoref{fig:N_opt_xrays}, we present the number of X-ray SNRs versus the number of optical SNRs. There is a trend in which galaxies with more optical SNRs tend to have more X-ray SNRs. This is expected, however, the galaxies M74, SMC, and LMC deviate from this trend, forming a 'parallel branch.' The biases affecting optical studies are even more pronounced since, in addition to the aforementioned factors, there is a significant variation in optical instruments and, consequently, in the sensitivity of the observations. This is also reflected in the fact that no correlation appears between the number of optical SNRs and SFR.

Despite the biases mentioned earlier, expanding the sample of galaxies would allow us to draw more accurate qualitative conclusions and better understand the factors influencing the detection of X-ray SNRs.

\begin{table*}
    \centering
    \caption{Information on galaxies with X-ray SNRs}
\begin{tabular}{llllll}
    \toprule
        Name & Distance & SFR & $\rm N_{optical}$ & $\rm N_{Xrays}$ & $\rm Lx_{min}$   \\
         &  (Mpc) &  (M$\rm \odot\, yr^{-1}$)  &  &  & $\rm (\times 10^{35}\, erg\, s^{-1})$  \\
    \hline
    NGC 7793 & 3.70\tablefootmark{1a} & 0.51\tablefootmark{1b, 1c} & 55\tablefootmark{1d} & 5\tablefootmark{*} & 82.7\tablefootmark{*}   \\
    LMC & 0.05\tablefootmark{2a}  & 0.35\tablefootmark{1b, 1c} & 49\tablefootmark{2b}  & 70\tablefootmark{2c, 2d} & 0.08\tablefootmark{2c, 2d}\\
    SMC & 0.06\tablefootmark{3a}  & 0.19\tablefootmark{1b, 1c} & 20\tablefootmark{3c}  & 17\tablefootmark{3c} & 0.05\tablefootmark{3c} \\
    M31 & 0.68\tablefootmark{4a} & 0.27\tablefootmark{1b, 1c} & 156\tablefootmark{4c} & 23\tablefootmark{4d} & 120\tablefootmark{4d} \\
    M33 & 0.80\tablefootmark{5a} & 0.45\tablefootmark{1b, 1c}  & 216\tablefootmark{5c, 5d}  & 112\tablefootmark{5e} &0.16\tablefootmark{5e}   \\
    NGC 4214 & 2.70\tablefootmark{6a}  & 0.12\tablefootmark{1b, 1c} & 92\tablefootmark{6b} & 7\tablefootmark{6b, 6c} & 25\tablefootmark{6c}\\
    NGC 2403 & 3.06\tablefootmark{7a} & 0.48\tablefootmark{1b, 1c} & 109\tablefootmark{6b} & 9\tablefootmark{6b, 6c} & 25\tablefootmark{6c}  \\
    NGC 3077 & 3.81\tablefootmark{8a} & 0.05\tablefootmark{1b, 1c} & 18{6b} & 1\tablefootmark{6b, 6c} & 400\tablefootmark{6c} \\ 
    NGC 6946 & 5.37\tablefootmark{9a} & 2.28\tablefootmark{1b, 1c} & 157\tablefootmark{9b} & 6\tablefootmark{9c} & 65\tablefootmark{9c} \\
    M101 & 7.2\tablefootmark{10a} & 1.41\tablefootmark{1b, 1c} & 93\tablefootmark{10d} & 7{\tablefootmark9c} & 19\tablefootmark{9c} \\ 
    M51 & 8.58\tablefootmark{11a} & 1.5\tablefootmark{1b, 1c}  & 179\tablefootmark{11b} & 55\tablefootmark{11b} & 11.1\tablefootmark{11b} \\ 
    M74 & 9.59\tablefootmark{12a} & 0.59\tablefootmark{1b, 1c} & 9\tablefootmark{12b} & 3\tablefootmark{12b} & 500\tablefootmark{12b} \\
    \hline
    \label{table:x_gal_info}
\end{tabular}

\tablefoot{\tablefoottext{1a}{\citet{2011ApJS..195...18R}}
\tablefoottext{1b}{\citet{2008ApJS..178..247K}}
\tablefoottext{1c}{\citet{2009ApJ...706..599L}}
\tablefoottext{1d}{\citet{2021MNRAS.507.6020K}}
\tablefoottext{*}{This work (\xmm{})}
\tablefoottext{2a}{\citet{2019Natur.567..200P}}
\tablefoottext{2b}{\citet{2021MNRAS.500.2336Y}}
\tablefoottext{2c}{\citet{2024A&A...692A.237Z}}
\tablefoottext{2d}{\citet{2025A&A...693L..15S}}
\tablefoottext{3a}{\citet{2013A&A...550A..70G}}
\tablefoottext{3c}{\citet{2019A&A...631A.127M}}
\tablefoottext{4a}{\citet{2011ApJ...743...19C}}
\tablefoottext{4c}{\citet{2014ApJ...786..130L}}
\tablefoottext{4d}{\citet{2012A&A...544A.144S}}
\tablefoottext{5a}{\citet{2002ApJ...565..959L}}
\tablefoottext{5c}{\citet{2010ApJS..187..495L}}
\tablefoottext{5d}{\citet{2014ApJ...793..134L}}
\tablefoottext{5e}{\citet{2017MNRAS.472..308G}}
\tablefoottext{6a}{\citet{2014A&A...566A..71L}}
\tablefoottext{6b}{\citet{2013MNRAS.429..189L}}
\tablefoottext{6c}{\citet{2010ApJ...725..842L}}
\tablefoottext{7a}{\citet{2008A&ARv..15..289T}}
\tablefoottext{8a}{\citet{2013AJ....146...86T}}
\tablefoottext{9a}{\citet{2014AJ....148..107R}}
\tablefoottext{9b}{\citet{2019ApJ...875...85L}}
\tablefoottext{9c}{\citet{2007AJ....133.1361P}}
\tablefoottext{10a}{\citet{2018ApJS..235...23S}}
\tablefoottext{10d}{\citet{1997ApJS..112...49M}}
\tablefoottext{11a}{\citet{2017AJ....154...51M}}
\tablefoottext{11b}{\citet{2021ApJ...908...80W}}
\tablefoottext{12a}{\citet{2017ApJ...834..174K}}
\tablefoottext{12b}{\citet{2010A&A...517A..91S}}
}

\end{table*}

\section{Conclusions} \label{sec:conclusions}
In this study, we analyze all archival {\textit{Chandra}} observations of the galaxy NGC 7793 to detect X-ray supernova remnants (SNRs), in addition to the single X-ray SNR previously known in this galaxy, primarily comparing with already known optical SNRs.
Our main findings are:
\begin{enumerate}

\item We detect five X-ray SNRs (X11, X15, X13, X25, X38), four of which are newly identified. These sources are considered X-ray SNRs primarily because they have optical SNR counterparts. Additional factors supporting this classification include:
\begin{enumerate}[a]
\item They exhibit soft X-ray emission, positioning them in the soft region of the color-color diagrams.
\item They show no significant short- or long-term variability.
\item Modeling of combined \xmm\ EPIC MOS spectra (1.1\,Ms) for X11 and X15 reveal soft emission typical of hot plasma (T >2.5$\times10^6$ K) with strong O\,VII, O\,VIII, and Ne\,IX K-shell lines.
\end{enumerate}

We also explore correlations between the X-ray and optical properties of SNRs. We find that all X-ray SNRs exhibit strong [\ion{O}{III}](5007\AA) emission and that there are positive correlations between X-ray luminosity and pre-shock density (calculated using the [\ion{S}{II}](6717)/[\ion{S}{II}](6731) ratio), as well as between color ($\rm log_{10}(S/H)$) and pre-shock density. However, although these trends are theoretically predicted, larger samples are required to confirm them with greater accuracy.

\item We report X-ray emission from SN 2008bk and refine its position.

\item We suggest two new candidate X-ray SNRs (X23 and X42). Although they have not been detected in optical wavelengths, we classify them as SNR candidates due to their soft, non-variable X-ray emission. Notably, for one of them (X42), the modeling of the combined \xmm\ EPIC MOS spectrum revealed soft emission from hot plasma (T > 2.5$\times10^6$ K) with strong K-shell emission lines of O\,VII, O\,VIII, and Ne\,IX, similar to the X-ray SNRs with optical SNR counterparts reported in this study.

\item Finally, we compare our results with those from studies of nearby galaxies where SNR populations have been examined in optical and X-ray wavelengths. We propose that distance, in combination with the limited sensitivity of the instruments, is a significant factor in not detecting as many SNRs as expected based on the SFR of the galaxies.

\end{enumerate}
With this study, we significantly increase the number of known X-ray SNRs in NGC 7793 from one to five. Additionally, this work contributes to the study of X-ray SNRs beyond the Local Group, where only a few have been identified. Systematic investigations of their properties require larger samples. Advances in optical identification techniques pave the way for detecting more optical SNRs, thereby increasing the likelihood of identifying additional X-ray counterparts.

\begin{acknowledgements}
We  thank the anonymous referee for the thorough review and the useful comments that helped to improve the clarity of the paper.
MK acknowledges financial support from MICINN (Spain) through the programme Juan de la Cierva-Incorporación [JC2022-049447-I]. MK and LG acknowledge financial support from AGAUR, CSIC, MCIN and AEI 10.13039/501100011033 under projects PID2023-151307NB-I00, PIE 20215AT016, CEX2020-001058-M, and 2021-SGR-01270. KA acknowledges support from \chandra{} grants GO3-24033B, GO0-21010X, and TM9-20001X, and \textit{JWST} grant JWST-GO-01905.002-A.  KA also made use of NASA’s Astrophysics Data System Bibliographic Services. 
NR is supported by the European Research Council (ERC) via the Consolidator Grant “MAGNESIA” (No. 817661) and the Proof of Concept "DeepSpacePulse" (No. 101189496), by the Catalan grant SGR2021-01269 (PI: Graber/Rea), the Spanish grant ID2023-153099NA-I00 (PI Coti Zelati), and by the program Unidad de Excelencia Maria de Maeztu CEX2020-001058-M.
CPG acknowledges financial support from the Secretary of Universities and Research (Government of Catalonia) and by the Horizon 2020 Research and Innovation Programme of the European Union under the Marie Skłodowska-Curie and the Beatriu de Pinós 2021 BP 00168 programme, from the Spanish Ministerio de Ciencia e Innovación (MCIN) and the Agencia Estatal de Investigación (AEI) 10.13039/501100011033 under the PID2020-115253GA-I00 HOSTFLOWS project, and the
program Unidad de Excelencia María de Maeztu CEX2020-001058-M. 

\end{acknowledgements}

\section*{Data Availability}
For this work we used archival data from {\it{Chandra}} and {\it{XMM-Newton}} telescopes.

\newpage


\clearpage
\begin{appendix} 
\onecolumn
\section{Properties of the \chandra{} detected sources}
\begin{table*}[htbp]
	\centering
\caption{Detected \chandra{}   sources in NGC\,7793 and their X-ray properties.}
\small
\begin{tabular}{@{}lllllllll@{}}
\hline
ID &\, \, \,  \, RA & \, \, \,  \, DEC & ncr broad ($\rm \times 10^{-5}$) & ncr soft ($\rm \times 10^{-5}$) & ncr medium ($\rm \times 10^{-5}$) & ncr hard ($\rm \times 10^{-5}$) & S/N$_{\rm broad}$ & S/N$_{\rm soft}$\\
 & \, \,  (J2000) & \, \,  \,  (J2000) &  0.5 - 7.0 keV & 0.5 - 1.2 keV & 1.2 - 2.0 keV & 2.0 - 7.0 keV & &\\
\hline
X1 & 23:57:35.0 & -32:27:54.3 & 97.66 $ \pm $ 13.86 & 28.14 $ \pm $ 7.20 & 29.87 $ \pm $ 7.40 & 39.65 $ \pm $ 9.25 & 5.88 & 2.87 \\
X2 & 23:57:36.5 & -32:38:44.9 & 85.34 $ \pm $ 13.70 & 35.93 $ \pm $ 8.27 & 26.80 $ \pm $ 7.37 & 22.62 $ \pm $ 8.06 & 5.08 & 3.29 \\
X3 & 23:57:36.8 & -32:29:26.6 & 70.93 $ \pm $ 12.33 & 24.23 $ \pm $ 6.48 & 28.46 $ \pm $ 7.47 & 18.24 $ \pm $ 7.37 & 4.64 & 2.70 \\
X4 & 23:57:38.7 & -32:30:45.5 & 8.88 $ \pm $ 5.24 & 1.73 $ \pm $ 1.73 & 3.06 $ \pm $ 3.24 & 4.08 $ \pm $ 3.74 & 1.04 & 0.34 \\
X5 & 23:57:39.0 & -32:39:08.6 & 274.33 $ \pm $ 30.56 & 40.42 $ \pm $ 11.67 & 94.32 $ \pm $ 17.82 & 139.59 $ \pm $ 21.91 & 7.89 & 2.43 \\
X6 & 23:57:42.7 & -32:37:17.2 & 46.87 $ \pm $ 9.88 & 7.46 $ \pm $ 4.38 & 20.39 $ \pm $ 5.89 & 19.01 $ \pm $ 6.61 & 3.69 & 1.00 \\
X7 & 23:57:43.2 & -32:33:56.6 & 32.00 $ \pm $ 7.89 & 8.66 $ \pm $ 3.87 & 10.39 $ \pm $ 4.24 & 12.96 $ \pm $ 5.41 & 3.03 & 1.29 \\
X8 & 23:57:43.5 & -32:29:28.4 & 43.85 $ \pm $ 9.56 & 5.51 $ \pm $ 3.60 & 11.41 $ \pm $ 4.63 & 26.93 $ \pm $ 7.54 & 3.54 & 0.82 \\
X9 & 23:57:43.8 & -32:36:34.7 & 273.53 $ \pm $ 22.31 & 84.05 $ \pm $ 12.38 & 101.50 $ \pm $ 13.33 & 87.97 $ \pm $ 12.92 & 11.04 & 5.67 \\
X10 & 23:57:43.9 & -32:34:21.8 & 44.27 $ \pm $ 9.14 & 18.69 $ \pm $ 5.64 & 21.26 $ \pm $ 6.18 & 4.31 $ \pm $ 3.69 & 3.78 & 2.29 \\
X11 & 23:57:44.0 & -32:34:40.9 & 60.62 $ \pm $ 11.49 & 62.44 $ \pm $ 11.41 & 0.00 $ \pm $ 0.00 & -1.82 $ \pm $ 1.29 & 4.20 & 4.39 \\
X12 & 23:57:44.1 & -32:28:43.3 & 374.64 $ \pm $ 25.92 & 139.19 $ \pm $ 15.74 & 141.26 $ \pm $ 15.69 & 94.19 $ \pm $ 13.34 & 13.27 & 7.72 \\
X13 & 23:57:45.8 & -32:35:01.5 & 12.05 $ \pm $ 6.17 & 9.14 $ \pm $ 4.69 & 0.97 $ \pm $ 2.31 & 1.94 $ \pm $ 3.27 & 1.23 & 1.12 \\
X14 & 23:57:46.8 & -32:36:07.7 & 954.90 $ \pm $ 44.50 & 335.37 $ \pm $ 26.40 & 313.73 $ \pm $ 25.38 & 305.80 $ \pm $ 25.28 & 20.33 & 11.59 \\
X15 & 23:57:47.3 & -32:35:23.5 & 82.62 $ \pm $ 13.37 & 79.60 $ \pm $ 13.01 & 4.09 $ \pm $ 2.89 & -1.07 $ \pm $ 1.07 & 5.09 & 5.03 \\
X16 & 23:57:48.2 & -32:36:57.8 & 73.17 $ \pm $ 11.86 & 45.35 $ \pm $ 9.08 & 13.06 $ \pm $ 5.26 & 14.76 $ \pm $ 5.53 & 5.05 & 3.92 \\
X17 & 23:57:48.5 & -32:28:51.5 & 63.10 $ \pm $ 11.34 & 22.80 $ \pm $ 6.56 & 28.71 $ \pm $ 7.17 & 11.58 $ \pm $ 5.84 & 4.47 & 2.47 \\
X18 & 23:57:48.7 & -32:32:33.7 & 68.55 $ \pm $ 12.21 & 37.83 $ \pm $ 8.96 & 28.61 $ \pm $ 7.65 & 2.10 $ \pm $ 3.21 & 4.53 & 3.17 \\
X19 & 23:57:49.9 & -32:35:28.3 & 2636.44 $ \pm $ 167.34 & 452.08 $ \pm $ 67.39 & 1092.18 $ \pm $ 108.31 & 1092.18 $ \pm $ 108.31 & 14.85 & 5.64 \\
X20 & 23:57:50.5 & -32:33:20.3 & 27.12 $ \pm $ 7.23 & 15.22 $ \pm $ 5.66 & 6.80 $ \pm $ 3.40 & 5.10 $ \pm $ 2.94 & 2.79 & 1.84 \\
X21 & 23:57:51.0 & -32:42:54.5 & 77.50 $ \pm $ 18.87 & 6.74 $ \pm $ 4.76 & 37.73 $ \pm $ 11.82 & 33.04 $ \pm $ 13.92 & 3.11 & 0.62 \\
X22 & 23:57:51.1 & -32:37:26.9 & 13341.46 $ \pm $ 165.91 & 4269.22 $ \pm $ 93.75 & 4103.96 $ \pm $ 92.00 & 4968.28 $ \pm $ 101.36 & 79.21 & 44.41 \\
X23 & 23:57:51.2 & -32:36:00.7 & 65.33 $ \pm $ 10.86 & 66.21 $ \pm $ 10.82 & 0.00 $ \pm $ 0.00 & -0.88 $ \pm $ 0.88 & 4.93 & 5.03 \\
X24 & 23:57:51.9 & -32:33:38.7 & 16.91 $ \pm $ 6.31 & 4.09 $ \pm $ 2.89 & 4.09 $ \pm $ 2.89 & 8.74 $ \pm $ 4.80 & 1.77 & 0.62 \\
X25 & 23:57:52.2 & -32:34:13.8 & 24.52 $ \pm $ 7.08 & 22.48 $ \pm $ 6.78 & 2.04 $ \pm $ 2.04 & 0.00 $ \pm $ 0.00 & 2.43 & 2.29 \\
X26 & 23:57:52.8 & -32:33:09.9 & 578.29 $ \pm $ 34.38 & 183.91 $ \pm $ 19.39 & 222.73 $ \pm $ 21.33 & 171.65 $ \pm $ 18.73 & 15.77 & 8.42 \\
X27 & 23:57:53.1 & -32:42:23.6 & 123.12 $ \pm $ 21.30 & 1.99 $ \pm $ 3.64 & 73.35 $ \pm $ 16.33 & 47.78 $ \pm $ 13.19 & 4.67 & 0.19 \\
X28 & 23:57:53.3 & -32:28:12.2 & 3438.21 $ \pm $ 77.72 & 1611.35 $ \pm $ 53.18 & 1085.82 $ \pm $ 43.54 & 741.05 $ \pm $ 36.28 & 43.03 & 29.14 \\
X29 & 23:57:54.1 & -32:31:35.0 & 35.64 $ \pm $ 8.74 & 14.30 $ \pm $ 5.41 & 7.03 $ \pm $ 4.24 & 14.30 $ \pm $ 5.41 & 3.04 & 1.66 \\
X30 & 23:57:54.9 & -32:39:55.9 & 151.89 $ \pm $ 16.39 & 59.90 $ \pm $ 10.24 & 59.48 $ \pm $ 10.05 & 32.51 $ \pm $ 7.92 & 8.12 & 4.76 \\
X31 & 23:57:56.2 & -32:36:33.1 & 60.41 $ \pm $ 10.43 & 19.57 $ \pm $ 5.94 & 24.67 $ \pm $ 6.63 & 16.17 $ \pm $ 5.44 & 4.70 & 2.29 \\
X32 & 23:57:56.2 & -32:31:37.1 & 15.58 $ \pm $ 5.19 & 5.19 $ \pm $ 3.00 & 3.46 $ \pm $ 2.45 & 6.92 $ \pm $ 3.46 & 1.99 & 0.86 \\
X33 & 23:57:56.4 & -32:34:45.9 & 23.98 $ \pm $ 6.75 & 3.40 $ \pm $ 2.40 & 13.59 $ \pm $ 4.81 & 6.99 $ \pm $ 4.09 & 2.58 & 0.62 \\
X34 & 23:57:56.5 & -32:36:00.3 & 149.94 $ \pm $ 17.85 & 55.56 $ \pm $ 10.94 & 59.26 $ \pm $ 11.00 & 35.12 $ \pm $ 8.83 & 7.32 & 4.03 \\
X35 & 23:57:56.7 & -32:27:50.0 & 34.97 $ \pm $ 15.53 & 7.33 $ \pm $ 7.54 & 12.14 $ \pm $ 7.01 & 15.50 $ \pm $ 11.63 & 1.51 & 0.48 \\
X36 & 23:57:57.5 & -32:31:49.5 & 22.50 $ \pm $ 6.24 & 3.46 $ \pm $ 2.45 & 8.66 $ \pm $ 3.87 & 10.39 $ \pm $ 4.24 & 2.57 & 0.62 \\
X37 & 23:57:57.9 & -32:26:46.8 & 31.24 $ \pm $ 9.96 & 20.97 $ \pm $ 6.64 & 9.08 $ \pm $ 4.34 & 1.20 $ \pm $ 6.03 & 2.31 & 2.20 \\
X38 & 23:57:59.2 & -32:36:06.6 & 20.43 $ \pm $ 6.46 & 12.26 $ \pm $ 5.01 & 6.13 $ \pm $ 3.54 & 2.04 $ \pm $ 2.04 & 2.14 & 1.48 \\
X39 & 23:57:59.9 & -32:32:41.1 & 185.27 $ \pm $ 20.39 & 80.38 $ \pm $ 13.52 & 61.30 $ \pm $ 11.19 & 43.59 $ \pm $ 10.37 & 8.13 & 4.99 \\
X40 & 23:57:59.9 & -32:33:20.2 & 18.39 $ \pm $ 11.00 & -2.04 $ \pm $ 8.91 & 10.22 $ \pm $ 4.57 & 10.22 $ \pm $ 4.57 & 1.58 & -0.23 \\
X41 & 23:58:00.2 & -32:33:26.0 & 94.34 $ \pm $ 13.21 & 76.17 $ \pm $ 11.48 & 19.04 $ \pm $ 5.74 & -0.87 $ \pm $ 3.12 & 6.13 & 5.56 \\
X42 & 23:58:00.4 & -32:34:54.8 & 124.65 $ \pm $ 15.96 & 114.43 $ \pm $ 15.29 & 10.22 $ \pm $ 4.57 & 0.00 $ \pm $ 0.00 & 6.74 & 6.41 \\
X43 & 23:58:00.7 & -32:32:39.5 & 25.5 $\pm$ 8.69 & 6.82$ \pm $4.75& 10.91$ \pm $5.56& 7.79 $\pm$4.70 & 2.03 & 0.79\\
X44 & 23:58:02.9 & -32:36:14.8 & 230.04 $ \pm $ 19.85 & 83.27 $ \pm $ 11.90 & 68.61 $ \pm $ 10.93 & 78.17 $ \pm $ 11.53 & 10.51 & 5.93 \\
X45 & 23:58:03.5 & -32:36:44.2 & 152.79 $\pm $ 17.90 & 69.48 $ \pm $ 11.92 & 45.74 $ \pm $ 9.88 & 37.57 $\pm$ 9.00 & 7.44 & 4.76 \\
X46 & 23:58:05.5 & -32:32:50.8 & 10.22 $\pm$ 5.60 & 2.55 $\pm$ 3.02 & 6.13 $\pm$ 4.21 &1.53 $\pm$ 2.11 & 1.08 & 0.36\\
X47 & 23:58:06.7 & -32:37:57.0 & 1558.32 $ \pm $ 56.89 & 631.49 $ \pm $ 36.18 & 527.25 $ \pm $ 33.01 & 399.57 $ \pm $ 28.95 & 26.24 & 16.34 \\
X48 & 23:58:07.4 & -32:26:06.5 & 53.15 $ \pm $ 12.27 & 13.04 $ \pm $ 5.35 & 23.72 $ \pm $ 7.39 & 16.39 $ \pm $ 8.21 & 3.35 & 1.56 \\
X49 & 23:58:07.8 & -32:37:15.9 & 22.46 $ \pm $ 7.07 & 11.11 $ \pm $ 4.56 & 7.71 $ \pm $ 3.88 & 3.65 $ \pm $ 3.75 & 2.26 & 1.51 \\
X50 & 23:58:07.9 & -32:36:15.1 & 49.04 $ \pm $ 10.01 & 18.39 $ \pm $ 6.13 & 12.26 $ \pm $ 5.01 & 18.39 $ \pm $ 6.13 & 3.83 & 1.99 \\
X51 & 23:58:08.4 & -32:38:48.0 & 237.13 $ \pm $ 20.81 & 29.25 $ \pm $ 7.27 & 83.33 $ \pm $ 12.23 & 124.54 $ \pm $ 15.18 & 10.18 & 2.98 \\
X52 & 23:58:08.8 & -32:34:03.8 & 3709.19 $ \pm $ 80.51 & 1751.79 $ \pm $ 55.27 & 1246.32 $ \pm $ 46.68 & 711.08 $ \pm $ 35.33 & 44.99 & 30.63 \\
X53 & 23:58:09.7 & -32:36:17.3 & 146.08 $ \pm $ 15.90 & 39.08 $ \pm $ 8.15 & 67.09 $ \pm $ 10.78 & 39.90 $ \pm $ 8.37 & 8.09 & 3.73 \\
X54 & 23:58:10.5 & -32:33:57.6 & 281.99 $ \pm $ 24.18 & 94.00 $ \pm $ 13.86 & 110.35 $ \pm $ 15.29 & 77.65 $ \pm $ 12.60 & 10.59 & 5.71 \\
X55 & 23:58:10.7 & -32:41:22.0 & 29.60 $ \pm $ 14.06 & -0.55 $ \pm $ 4.06 & 6.18 $ \pm $ 6.26 & 23.98 $ \pm $ 11.91 & 1.43 & -0.05 \\
X56 & 23:58:11.3 & -32:33:26.3 & 59.48 $ \pm $ 10.05 & 18.69 $ \pm $ 5.64 & 28.89 $ \pm $ 7.01 & 11.90 $ \pm $ 4.50 & 4.84 & 2.29 \\
X57 & 23:58:14.2 & -32:29:03.5 & 45.61 $ \pm $ 9.71 & 15.58 $ \pm $ 5.19 & 22.13 $ \pm $ 6.59 & 7.90 $ \pm $ 4.89 & 3.64 & 1.99 \\
X58 & 23:58:15.8 & -32:31:09.2 & 52.13 $ \pm $ 9.70 & 17.31 $ \pm $ 5.47 & 24.23 $ \pm $ 6.48 & 10.58 $ \pm $ 4.71 & 4.29 & 2.14 \\
\hline
 \end{tabular}
 \label{table:source_info}
 \end{table*}
 
\section{Colors of the X-ray detected sources}
\begin{table}[htbp]
	\centering
\caption{Colors and the corresponding OBSIDs of the detected sources}
\tiny
\vspace{-0.2cm}
\begin{tabular}{@{}llcccl@{}}
\toprule
ID &OBSID &  log$_{10}$(S/H) & log$_{10}$(S/M) & log$_{10}$(M/H) & Other Surveys\\
\hline
 X1 & 13439  & 	-0.07 $^{+ 0.16 } _{- 0.15 }$ & 0.06 $^{+ 0.15 } _{- 0.16 }$ & -0.14 $^{+ 0.16 } _{- 0.13 }$ \\[0.55pt]
 X2 & 14231  &	0.24 $^{+ 0.22 } _{- 0.16 }$ & 0.20 $^{+ 0.14 } _{- 0.18 }$ & 0.09 $^{+ 0.20 } _{- 0.21 }$ \\[0.55pt]
 X3 & 13439  & 	0.26 $^{+ 0.21 } _{- 0.24 }$ & 0.03 $^{+ 0.16 } _{- 0.16 }$ & 0.21 $^{+ 0.23 } _{- 0.22 }$ \\[0.55pt]
X4 & 13439 &  	-0.20 $^{+ 0.74 } _{- 0.73 }$ & -0.08 $^{+ 0.74 } _{- 0.80 }$ & -0.14 $^{+ 0.84 } _{- 0.72 }$ \\[0.55pt]
 X5 & 23266 & 	-0.38 $^{+ 0.14 } _{- 0.15 }$ & -0.25 $^{+ 0.13 } _{- 0.17 }$ & -0.13 $^{+ 0.11 } _{- 0.10 }$ \\[0.55pt]
X6 & 14231 & 	-0.39 $^{+ 0.37 } _{- 0.31 }$ & -0.38 $^{+ 0.30 } _{- 0.34 }$ & 0.04 $^{+ 0.21 } _{- 0.20 }$ \\[0.55pt]
X7 & 13439 & -0.05 $^{+ 0.31 } _{- 0.27 }$ & 0.03 $^{+ 0.27 } _{- 0.28 }$ & -0.04 $^{+ 0.26 } _{- 0.30 }$ \\[0.55pt]
X8 & 13439 & -0.66 $^{+ 0.36 } _{- 0.34 }$ & -0.20 $^{+ 0.29 } _{- 0.49 }$ & -0.32 $^{+ 0.19 } _{- 0.26 }$ \\[0.55pt]
X9 & 14231 &  	0.00 $^{+ 0.09 } _{- 0.09 }$ & -0.07 $^{+ 0.09 } _{- 0.09 }$ & 0.08 $^{+ 0.08 } _{- 0.09 }$ & \tiny{J235743.8-323633$^{(p)}$}\\[0.55pt]
X10 & 14231 & 	0.57 $^{+ 0.57 } _{- 0.32 }$ & 0.00 $^{+ 0.16 } _{- 0.21 }$ & 0.63 $^{+ 0.51 } _{- 0.38 }$ \\[0.55pt]
X11 & 3954 &  1.84 $^{+ 0.92 } _{- 0.60 }$ &  2.00 $^{+ 0.76 } _{- 0.76 }$ & -0.13 $^{+ 1.32 } _{- 1.03 }$ & \tiny{7793\_24$^{(k1)}$, SNR\_91$^{(k2)}$, J235743.9-323441$^{(mn)}$}\\[0.55pt]
X12 & 13439 & 	0.18 $^{+ 0.07 } _{- 0.08 }$ & 0.01 $^{+ 0.06 } _{- 0.07 }$ & 0.19 $^{+ 0.07 } _{- 0.09 }$ & \tiny{P5$^{(r)}$}\\[0.55pt]
X13 & 3954 &  	0.47 $^{+ 0.78 } _{- 0.56 }$ & 0.73 $^{+ 0.74 } _{- 0.66 }$ & -0.26 $^{+ 1.04 } _{- 0.90 }$ & \tiny{7793\_23$^{(k1)}$, SNR\_126$^{(k2)}$ }\\[0.55pt]
X14 & 3954 & 0.03 $^{+ 0.05 } _{- 0.05 }$ & 0.03 $^{+ 0.05 } _{- 0.05 }$& 0.00 $^{+ 0.04 } _{- 0.06 }$ & \tiny{J235746.7-323607$^{(p)}$, SNR R3$^{(pr)}$} \\[0.55pt]
X15 & 3954 &1.88 $^{+ 0.96 } _{- 0.56 }$ & 1.22 $^{+ 0.41 } _{- 0.31 }$ & 0.81 $^{+ 0.86 } _{- 0.93 }$ & \tiny{S11$^{(b97, pr)}$, 7793\_22$^{(k1)}$, J235747.2-323523$^{(p)}$, SNR\_136$^{(k2)}$}\\[0.55pt]
X16 & 14231 &  	0.49 $^{+ 0.19 } _{- 0.20 }$ & 0.55 $^{+ 0.19 } _{- 0.22 }$ & -0.05 $^{+ 0.25 } _{- 0.26 }$ \\[0.55pt]
X17 & 13439 &  	0.26 $^{+ 0.33 } _{- 0.24 }$ & -0.08 $^{+ 0.15 } _{- 0.18 }$ & 0.39 $^{+ 0.28 } _{- 0.27 }$ \\[0.55pt]
X18 & 3954 & 1.02 $^{+ 0.73 } _{- 0.41 }$  & 0.14 $^{+ 0.15 } _{- 0.16 }$ & 1.08 $^{+ 0.53 } _{- 0.60 }$ & \tiny{H18$^{(h)}$, 0574-1250312$^{(u)}$, P6$^{(r)}$, J235748.6-323234$^{(p)}$}\\[0.55pt]
X19 & 27481 & 	-0.38 $^{+ 0.09 } _{- 0.07 }$ & -0.39 $^{+ 0.09 } _{- 0.07 }$ & 0.01 $^{+ 0.06 } _{- 0.06 }$ & \tiny{J235749.9-323526$^{(b)}$}\\[0.55pt]
X20 & 14231 &  0.49 $^{+ 0.40 } _{- 0.36 }$ & 0.39 $^{+ 0.33 } _{- 0.35 }$ & 0.21 $^{+ 0.38 } _{- 0.53 }$ & \tiny{SN 2008bk$^{(v)}$}\\[0.55pt]
X21 & 23266 & 	-0.58 $^{+ 0.42 } _{- 0.41 }$ & -0.62 $^{+ 0.32 } _{- 0.42 }$ & 0.05 $^{+ 0.27 } _{- 0.22 }$ & \tiny{P21$^{(r)}$}\\[0.55pt]
X22 & 3954 &  	-0.07 $^{+ 0.01 } _{- 0.01 }$ & 0.02 $^{+ 0.01 } _{- 0.01 }$ & -0.09 $^{+ 0.01 } _{- 0.01 }$ & \tiny{P13$^{(r)}$, J235750.9-323726$^{(p)}$}\\[0.55pt]
X23 & 14231 & 	1.89 $^{+ 0.96 } _{- 0.56 }$ & 2.12 $^{+ 0.75 } _{- 0.77 }$ & 0.05 $^{+ 1.14 } _{- 1.21 }$ \\[0.55pt]
X24 & 3954 &  	-0.29 $^{+ 0.48 } _{- 0.55 }$ & -0.01 $^{+ 0.59 } _{- 0.58 }$ & -0.28 $^{+ 0.48 } _{- 0.55 }$ \\[0.55pt]
X25 & 3954 &1.34 $^{+ 0.96 } _{- 0.59 }$ & 1.00 $^{+ 0.55 } _{- 0.53 }$ & 0.34 $^{+ 1.15 } _{- 0.89 }$ & \tiny{J235752.2-323413$^{(p)}$, J235752.2-323413$^{(mn)}$, 7793\_5$^{(k1)}$}\\[0.55pt]
X26 & 3954 &	0.03 $^{+ 0.06 } _{- 0.07 }$ & -0.09 $^{+ 0.07 } _{- 0.05 }$ &  0.11 $^{+ 0.07 } _{- 0.06 }$ & \tiny{P7$^{(r)}$, 0574-1250339$^{(u)}$, J235752.7-323309$^{(p)}$}\\[0.55pt]
X27 & 23266 & 	-1.00 $^{+ 0.41 } _{- 0.81 }$ & -1.27 $^{+ 0.46 } _{- 0.74 }$ & 0.19 $^{+ 0.16 } _{- 0.15 }$ \\[0.55pt]
X28 & 13439 &  	0.34 $^{+ 0.02 } _{- 0.03 }$ & 0.17 $^{+ 0.02 } _{- 0.02 }$ & 0.17 $^{+ 0.03 } _{- 0.03 }$ & \tiny{P4$^{(r)}$}\\[0.55pt]
X29 & 3954 & 	0.01 $^{+ 0.27 } _{- 0.22 }$ & 0.31 $^{+ 0.37 } _{- 0.35 }$ & -0.29 $^{+ 0.35 } _{- 0.35 }$ \\[0.55pt]
X30 & 14231 & 	0.23 $^{+ 0.13 } _{- 0.13 }$ & 0.01 $^{+ 0.10 } _{- 0.11 }$ & 0.23 $^{+ 0.13 } _{- 0.13 }$ \\[0.55pt]
X31 & 14231 &  	0.07 $^{+ 0.19 } _{- 0.21 }$ & -0.10 $^{+ 0.18 } _{- 0.18 }$ & 0.15 $^{+ 0.20 } _{- 0.18 }$ \\[0.55pt]
X32 & 13439 & 	-0.17 $^{+ 0.56 } _{- 0.48 }$ & 0.13 $^{+ 0.69 } _{- 0.59 }$ & -0.25 $^{+ 0.55 } _{- 0.64 }$ \\[0.55pt]
X33 & 14231 & 	-0.27 $^{+ 0.53 } _{- 0.58 }$ & -0.53 $^{+ 0.35 } _{- 0.52 }$ & 0.31 $^{+ 0.34 } _{- 0.39 }$ & \tiny{J235756.3-323444$^{(p)}$}\\[0.55pt]
X34 & 3954 & 	0.17 $^{+ 0.15 } _{- 0.14 }$ & -0.04 $^{+ 0.12 } _{- 0.11 }$ & 0.20 $^{+ 0.15 } _{- 0.13 }$ & \tiny{J235756.4-323559$^{(p)}$}\\[0.55pt]
X35 & 14378 & 	-0.25 $^{+ 0.68 } _{- 0.76 }$ & -0.13 $^{+ 0.49 } _{- 0.70 }$ & -0.11 $^{+ 0.55 } _{- 0.47 }$ \\[0.55pt]
X36 & 13439 & 	-0.44 $^{+ 0.45 } _{- 0.55 }$ & -0.40 $^{+ 0.52 } _{- 0.53 }$ & -0.06 $^{+ 0.34 } _{- 0.35 }$ \\[0.55pt]
X37 & 13439 & 	0.70 $^{+ 0.77 } _{- 0.52 }$ & 0.37 $^{+ 0.25 } _{- 0.28 }$ & 0.25 $^{+ 0.92 } _{- 0.47 }$ \\[0.55pt]
X38 & 3954 & 	0.69 $^{+ 0.58 } _{- 0.52 }$ & 0.24 $^{+ 0.41 } _{- 0.31 }$ & 0.36 $^{+ 0.70 } _{- 0.56 }$ & \tiny{S24$^{(b98)}$, 7793\_21$^{(k1)}$}\\[0.55pt]
X39 & 3954 & 0.26 $^{+ 0.15 } _{- 0.12 }$ & 0.12 $^{+ 0.11 } _{- 0.12 }$ & 0.17 $^{+ 0.12 } _{- 0.15 }$ & \tiny{J235759.8-323240$^{(p)}$}\\[0.55pt]
X40 & 3954 & 	-0.38 $^{+ 0.68 } _{- 0.99 }$ & -0.38 $^{+ 0.68 } _{- 0.99 }$ & 0.05 $^{+ 0.54 } _{- 0.66 }$ \\[0.55pt]
X41 & 13439 &  	1.57 $^{+ 0.95 } _{- 0.40 }$ & 0.60 $^{+ 0.17 } _{- 0.15 }$ & 1.04 $^{+ 0.88 } _{- 0.51 }$ & \tiny{P8$^{(r)}$, J235800.1-323325$^{(p)}$}\\[0.55pt]
X42 & 3954 &	2.16 $^{+ 0.85 } _{- 0.66 }$ & 1.04 $^{+ 0.28 } _{- 0.22 }$ & 1.01 $^{+ 0.98 } _{- 0.70 }$ & \tiny{J235800.3-323455$^{(p)}$} \\[0.55pt]
X43 & 14231 & 	0.01 $^{+ 0.10 } _{- 0.08 }$ & 0.09 $^{+ 0.09 } _{- 0.09 }$ & -0.07 $^{+ 0.09 } _{- 0.10 }$ & \tiny{J235802.8-323614$^{(p)}$} \\[0.55pt]
X44 & 14231 &  0.01 $^{+ 0.10 } _{- 0.08 }$ & 0.09 $^{+ 0.09 } _{- 0.09 }$ & -0.07 $^{+ 0.09} _{- 0.10 }$ \\[0.55pt]

X45 & 3954  & 	0.26 $^{+ 0.14 } _{- 0.12 }$ & 0.19 $^{+ 0.12 } _{- 0.12 }$ & 0.07 $^{+ 0.15 } _{- 0.14 }$ & \tiny{J235803.5-323643$^{(p)}$}\\[0.55pt]
X46 & 3954 &  	0.25 $^{+ 0.04 } _{- 0.04 }$ & 0.13 $^{+ 0.03 } _{- 0.04 }$ & 0.12 $^{+ 0.04 } _{- 0.04 }$ & \tiny{J235806.6-323757$^{(m)}$} \\[0.55pt]
X47 & 13439 & 	0.02 $^{+ 0.34 } _{- 0.32 }$ & -0.15 $^{+ 0.21 } _{- 0.26 }$ & 0.13 $^{+ 0.37 } _{- 0.23 }$ \\[0.55pt]
X48 & 13439 &  0.02 $^{+ 0.34 } _{- 0.32 }$ & -0.16 $^{+ 0.21} _{- 0.26 }$ & 0.13 $^{+ 0.37 } _{- 0.23 }$ \\[0.55pt]

X49 & 14231 & 	0.42 $^{+ 0.62 } _{- 0.44 }$ & 0.19 $^{+ 0.29 } _{- 0.33 }$ & 0.26 $^{+ 0.62 } _{- 0.53 }$ \\[0.55pt]
X50 & 3954 & 	0.04 $^{+ 0.19 } _{- 0.24 }$ & 0.20 $^{+ 0.24 } _{- 0.25 }$ & -0.13 $^{+ 0.20 } _{- 0.29 }$ & \tiny{J235807.8-323614$^{(p)}$}\\[0.55pt]
X51 & 14231 &  	-0.53 $^{+ 0.11 } _{- 0.13 }$ & -0.36 $^{+ 0.11 } _{- 0.14 }$ & -0.15 $^{+ 0.07 } _{- 0.10 }$ \\[0.55pt]
X52 & 13439 & 	0.41 $^{+ 0.02 } _{- 0.03 }$ & 0.16 $^{+ 0.02 } _{- 0.02 }$ & 0.25 $^{+ 0.03 } _{- 0.03 }$ & \tiny{P9$^{(r)}$, J235808.7-323403$^{(p)}$}\\[0.55pt]
X53 & 14231 & 	0.03 $^{+ 0.14 } _{- 0.12 }$ & -0.17 $^{+ 0.10 } _{- 0.13 }$ & 0.22 $^{+ 0.12 } _{- 0.11 }$ & \tiny{J235807.8-323614$^{(p)}$} \\[0.55pt]
X54 & 3954 & 	0.07 $^{+ 0.11 } _{- 0.09 }$ & -0.08 $^{+ 0.10 } _{- 0.08 }$ & 0.14 $^{+ 0.10 } _{- 0.09 }$ & \tiny{J235810.4-323357$^{(p)}$} \\[0.55pt]
X55 & 23266 & -0.83 $^{+ 0.55 } _{- 1.00 }$ & -0.39 $^{+ 0.89 } _{- 1.01 }$ & -0.56 $^{+ 0.61 } _{- 0.56 }$ \\[0.55pt]
X56 & 14231 & 	0.20 $^{+ 0.26 } _{- 0.19 }$ & -0.16 $^{+ 0.16 } _{- 0.19 }$ & 0.35 $^{+ 0.25 } _{- 0.16 }$ \\[0.55pt]
X57 & 13439 &  0.51 $^{+ 0.35 } _{- 0.35 }$ & 0.03 $^{+ 0.16 } _{- 0.23 }$ & 0.43 $^{+ 0.42 } _{- 0.27 }$ \\[0.55pt]
X58 & 13439 &  	0.38 $^{+ 0.25 } _{- 0.25 }$ & -0.04 $^{+ 0.19 } _{- 0.18 }$ & 0.41 $^{+ 0.22 } _{- 0.25 }$ \\[0.55pt]
\bottomrule
\end{tabular}
\vspace{-0.2cm}
\tiny
\tablefoot{$^{(p)}$\citet{2011AAS...21725633P}; $^{(k1)}$\citet{2021MNRAS.507.6020K}; $^{(k2)}$\citet{2024MNRAS.530.1078K}; $^{(mn)}$\citet{2012MNRAS.419.2095M}; $^{(r)}$\citet{1999A&A...341....8R}; $^{(pr)}$\citet{2002ApJ...565..966P}; $^{(b97)}$\citet{Blair1997}; $^{(h)}$\citet{1969ApJS...18...73H}; $^{(u)}$USNO: \citet{2003AJ....125..984M}; $^{(m)}$\citet{2011AN....332..367M}; $^{(b)}$\citet{2023ApJ...951...51B}; $^{(v)}$\citet{2013AJ....146...24V}}
\label{table:fluxes}
\end{table}

\twocolumn
\section{X-ray sources unrelated to Supernova Remnants} \label{sec:comparison_ap}
In this section, we compare the detected sources that are unrelated to SNRs with those reported in previous studies. Source X18 had been misclassified as an H II region (H18; \citealt{1969ApJS...18...73H}), but as mentioned in \citet{2011AJ....142...20P} (CXOUJ235748.6-323234) it is a foreground star (USNO 0574-1250312; \citealt{2003AJ....125..984M}). This source has been earlier detected and labeled as P6 in \citet{1999A&A...341....8R} by the ROSAT PSPC).  Source X26 is also a foreground star (USNO 0574-1250339; \citealt{2003AJ....125..984M}) and has been presented in  \citealt{2011AJ....142...20P} as CXOU J235752.7-323309 and in  \citet{1999A&A...341....8R} as P7.

Source X41 (P8 in \citealt{1999A&A...341....8R}; CXOU J235800.1-323325 in \citealt{2011AJ....142...20P}) was first misclassified as SNR (S26 in \citealt{Blair1997}) but it is a microquasar (\citealt{2010Natur.466..209P}, \citealt{2010MNRAS.409..541S}). According to \citet{2011AJ....142...20P} the spatial extent in the X-ray, optical and radio wavelengths is similar. There is a super-bubble around this source (\citealt{2021MNRAS.507.6020K}) which is created by the microquasar \citep{2012MNRAS.427..956D}. X40 is also part of this bubble.

Source X52 is reported as CXOU J235808.7-323403 in \citet{2011AJ....142...20P} and as P9 in \citet{1999A&A...341....8R}, both in X-rays. According to the latter it presents a soft spectrum, it is variable and highly absorbed. X22 is assumed to be either a background galaxy or a black-hole X-ray binary (CXOU J235750.9-323726 in \citealt{2011AJ....142...20P}; P13 in \citealt{1999A&A...341....8R}). \citet{2011AN....332..367M}  identified a star with a  V magnitude of 20.5 in the optical and suggested it as companion of this X-ray source. X46 is reported as variable and it is probably a X-ray binary (CXOU J235806.6-323757 in \citealt{2011AN....332..367M}). Source X19 coincides with the nuclear position of the galaxy (J235749.9-323526 in \citealt{2023ApJ...951...51B}).

The  sources X33 (CXOU J235756.3-323444), X34 (CXOU J235756.4-323559), X39 (CXOUJ235759.8-323240), X42 (CXOU J235800.3-323455), X43 (CXOU J235802.8-323614),  X45 (CXOU J235803.5-323643),  X50 (CXOU J235807.8-323614),  X53 (CXOU J235809.6-323617), and X54 (CXOU J235810.4-323357) are presented in \citet{2011AJ....142...20P} with their names indicated in the parenthesis. X12, X21, and X28 have been detected before by \citet{1999A&A...341....8R} as P5, P21, and P4 respectively. X12 is reported as a faint source, X21 as a variable object associated to the galaxy, and X28 as a stellar object associated with the galaxy.
 



\end{appendix}
\label{lastpage}
\end{document}